\documentclass[prl,twocolumn,superscriptaddress,groupedaddress]{revtex4}  

\usepackage{amssymb}
\usepackage{amsmath}
\usepackage{esint}
\usepackage{epsfig}
\usepackage{epstopdf}
\usepackage{bm}
\usepackage{graphicx,epsfig}
\usepackage{mathrsfs}
\usepackage{dcolumn}
\usepackage{color}
\usepackage{natbib}
\usepackage{CJK}
\usepackage{miniplot}
\usepackage{subfigure}
\usepackage{minitoc}
\usepackage{blindtext}
\usepackage[utf8]{inputenc}

\def\be{\begin{equation}}
\def\ee{\end{equation}}
\def\bea{\begin{eqnarray}}
\def\eea{\end{eqnarray}}
\def\la{\langle}
\def\ra{\rangle}

\allowdisplaybreaks[2]

\begin{document}

\title{Topological phase transitions in superradiance lattices}
\author{Da-Wei Wang}
\email{cuhkwdw@gmail.com}
\affiliation{Department of Physics and Centre for Quantum Coherence, The Chinese University of Hong Kong, Hong Kong, China}
\affiliation{Institute for Quantum Science and Engineering and Department of Physics and Astronomy, Texas A$\&$M University, College Station, TX 77843, USA}
\author{Han Cai}
\affiliation{Institute for Quantum Science and Engineering and Department of Physics and Astronomy, Texas A$\&$M University, College Station, TX 77843, USA}
\author{Luqi Yuan}
\affiliation{Institute for Quantum Science and Engineering and Department of Physics and Astronomy, Texas A$\&$M University, College Station, TX 77843, USA}
\affiliation{Department of Electrical Engineering, and Ginzton
Laboratory, Stanford University, Stanford, CA 94305, USA}
\author{Shi-Yao Zhu}
\affiliation{Beijing Computational Science Research Centre, Beijing 100084, China}
\affiliation{Synergetic Innovation Center of Quantum Information and Quantum Physics, University of Science and Technology of China, Hefei, Anhui 230026, China}
\author{Ren-Bao Liu}
\email{rbliu@phy.cuhk.edu.hk}
\affiliation{Department of Physics and Centre for Quantum Coherence, The Chinese University of Hong Kong, Hong Kong, China}

\date{\today }

\begin{abstract}
Topological phases of matters are of fundamental interest and have promising applications. Fascinating topological properties of light have been unveiled in classical optical materials. However, the manifestation of topological physics in quantum optics has not been discovered. Here we study the topological phases in a two-dimensional momentum-space superradiance lattice composed of timed Dicke states (TDS) in electromagnetically induced transparency (EIT). By periodically modulating the three EIT coupling fields, we can create a Haldane model with in-situ tunable topological properties, which manifest themselves in the contrast between diffraction signals emitted by superradiant TDS. The topological superradiance lattices provide a controllable platform for simulating exotic phenomena in condensed matter physics and offer a basis of topological quantum optics and novel photonic devices.
\end{abstract}

\maketitle

\noindent The quantum Hall effect (QHE) \cite{Klitzing1980} reveals a new class of matter phases, topological insulators (TIs), which have been extensively studied in solid-state materials \cite{Haldane1988,Kane2005,Bernevig2006,Konig2007,Chang2013,Qi2009} and recently in photonic structures \cite{Haldane2008,HafeziM2013,Rechtsman2013,Khanikaev2013,Nalitov2015}, time-periodic systems \cite{Oka2009,Inoue2010,Kitagawa2010,Lindner2011} and optical lattices of cold atoms \cite{Jotzu2014}.
The first TI is the Haldane model \cite{Haldane1988} proposed in 1988, which shows that the QHE is an intrinsic topological property of the energy bands. It inspired the discovery of the quantum spin Hall effect \cite{Kane2005,Bernevig2006,Konig2007} and topological superconductors \cite{Qi2009}. The Haldane model consists of a honeycomb tight-binding lattice with complex next-nearest-neighbour (NNN) hopping, which breaks the time-reversal symmetry and induces an energy gap between two bands that have opposite Chern numbers. Inversion symmetry breaking by an on-site potential can also open band gaps. The interplay of these two types of symmetry breaking leads to transitions between phases with Chern numbers 0 and $\pm 1$. 

Notwithstanding its foundational role in topological condensed matter physics, the Haldane model has never been realized in solid-state systems. Floquet modulation of a circular polarised light in graphene was proposed to induce the complex NNN tunnelling \cite{Oka2009,Inoue2010,Kitagawa2010}. However, the required light is soft X-ray which would directly excite electrons and be absorbed. The Haldane model of cold atoms in optical lattices \cite{Jotzu2014} were recently realised in experiments.

\begin{figure} 
\epsfig{figure=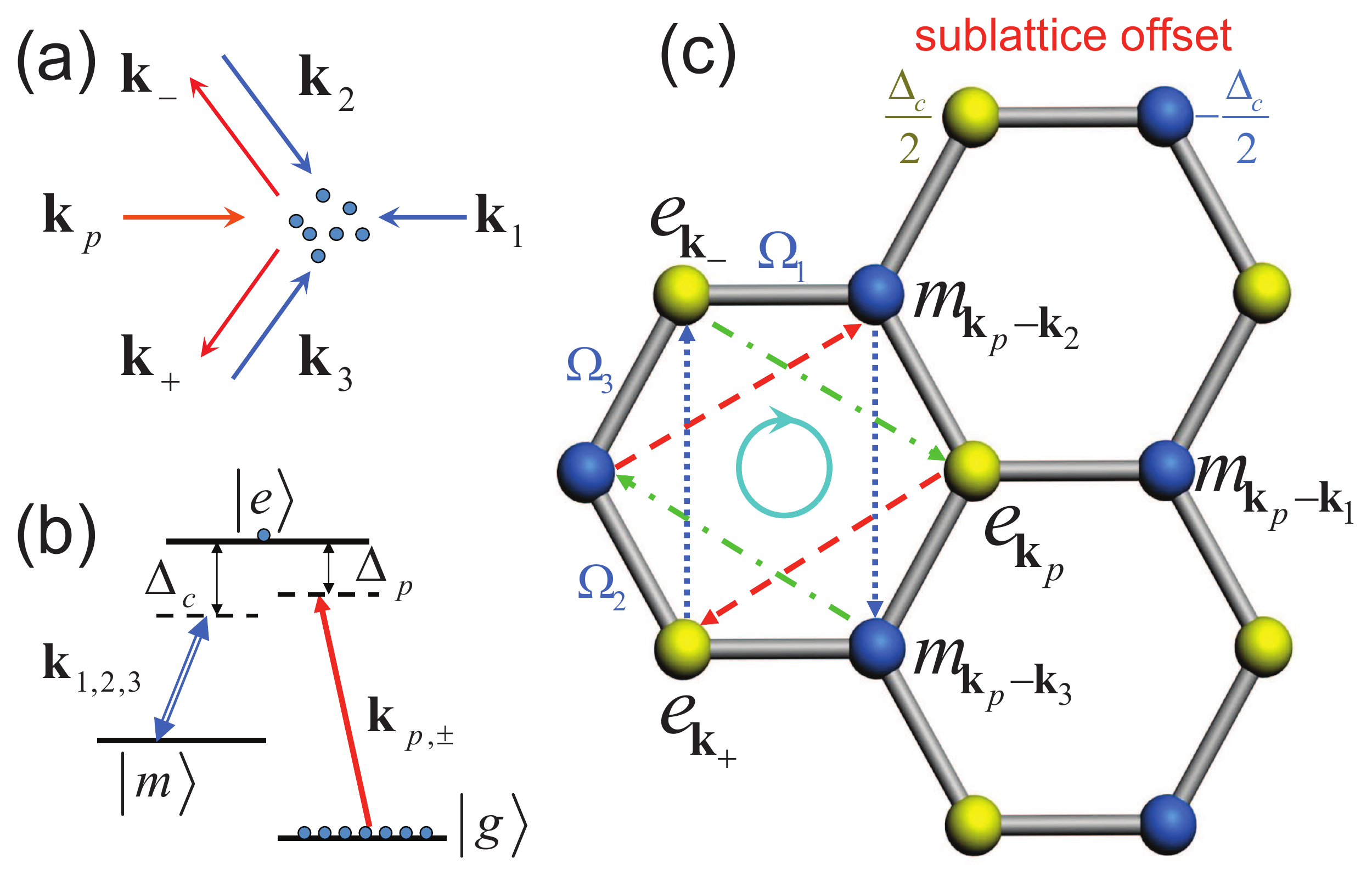, angle=0, width=0.5\textwidth}   
  \caption{{Realisation of the Haldane model in a superradiance lattice of timed Dicke States in electromagnetically induced transparency.} (a) Schematic configuration of the probe field $\mathbf{k}_p$, EIT coupling fields along $\mathbf{k}_{1/2/3}$ and diffraction fields along $\mathbf{k}_+\equiv\mathbf{k}_p+\mathbf{k}_1-\mathbf{k}_3$ and $\mathbf{k}_-\equiv\mathbf{k}_p+\mathbf{k}_1-\mathbf{k}_2$. ({b}) The energy level diagram of the EIT coupling, probe and scattering fields. ({c}) Honeycomb structure of the SL. $|e_\mathbf{k}\ra$ and $|m_{\mathbf{k}}\ra$ correspond to the two sublattices. The NNN hopping $\Omega_{31}$ (red solid arrow), $\Omega_{23}$ (blue dot arrow) and $\Omega_{12}$ (green dash dot arrow) enclose a nonzero effective magnetic flux in momentum space. The arrow on the circle denotes the direction of excitation current when the Chern number $C=1$.} 
  \label{graphene} 
\end{figure}

Our recent study shows that Scully's timed Dicke states (TDS) \cite{Scully2006} can form a superradiance lattice (SL) in momentum space \cite{Wang2015}. Here we propose a quantum optics realisation of an in-situ tunable Haldane model using two-dimensional SLs in a simple electromagnetically induced transparency (EIT) configuration \cite{Boller1991}. Since Dicke's seminal paper in 1954 \cite{Dicke1954}, superradiance has been an important topic in quantum optics. A single photon with wave vector $\mathbf{k}_p$ can excite a spatially extended $N$-atom ensemble from the ground state $|G\rangle=|g_1,g_2,\ldots, g_N\rangle$ to the TDS \cite{Scully2006}
\begin{equation}
|e_{\mathbf{k}_p}\ra=\frac{1}{\sqrt{N}}\sum\limits_{j=1}^{N}e^{i\mathbf{k}_p\cdot \mathbf{r}_j}|g_1, g_2,...,e_j,...,g_N\ra,
\label{tds}
\end{equation} 
where $e_i$ and $g_i$ are the excited and ground states of the $i$th atom at position $\mathbf{r}_i$, respectively. The TDS stores a light momentum $\hbar\mathbf{k}_p$ via phase correlations between atoms excited at different positions. This momentum can be transferred back to a single photon via directional emission \cite{Scully2006}. By coupling $|e\ra$ to another state $|m\ra$ with three coherent plane wave fields, we construct a honeycomb SL of TDS in momentum space \cite{Wang2015}, as shown in Fig.\ref{graphene}. The SL Hamiltonian with rotating-wave approximation is (see Supplementary Information)
\begin{equation}
\begin{aligned}
H=&\frac{\hbar\Delta_c}{2}\sum\limits_{\mathbf{k}}(|e_\mathbf{k}\ra\la e_\mathbf{k}|-|m_{\mathbf{k}-\mathbf{k}_1}\ra\la m_{\mathbf{k}-\mathbf{k}_1}|)\\
&-\sum\limits_{\mathbf{k}}\sum\limits_{l=1}^{3}\hbar\Omega_l [|e_{\mathbf{k}}\ra\la m_{\mathbf{k}-\mathbf{k}_l}|+h.c.],
\label{Hh}
\end{aligned}
\end{equation}
where $\mathbf{k}=\mathbf{k}_p+r(\mathbf{k}_2-\mathbf{k}_1)+s(\mathbf{k}_3-\mathbf{k}_2)$ with integers $r$ and $s$, $\Omega_l$ is the Rabi frequency of the coupling field along $\mathbf{k}_l$, $|m_\mathbf{k}\ra$ are defined the same as in Eq.(\ref{tds}) but for $|m\ra$ states, and $\Delta_c=\omega_{em}-\nu_c$ is the detuning of the transition frequency $\omega_{em}$ between $|e\ra$ and $|m\ra$ from the angular frequency $\nu_c$ of the EIT coupling fields. The states $|e_{\mathbf{k}}\rangle$ and $|m_{\mathbf{k}-\mathbf{k}_1}\rangle$ form two sublattices of a honeycomb structure in momentum space [see Fig. \ref{graphene}(c)]. $\Omega_l$ causes the nearest-neighbour hopping. As we will show below, periodically modulating $\Omega_l$ can introduce NNN hopping with controllable phases. On the other hand, the detuning $\Delta_c$ causes an energy offset between the two sublattices, breaking the inversion symmetry. Thus three Haldane phases with Chern numbers $\pm 1$ and $0$ can be in-situ realised. 

\begin{figure} 
\epsfig{figure=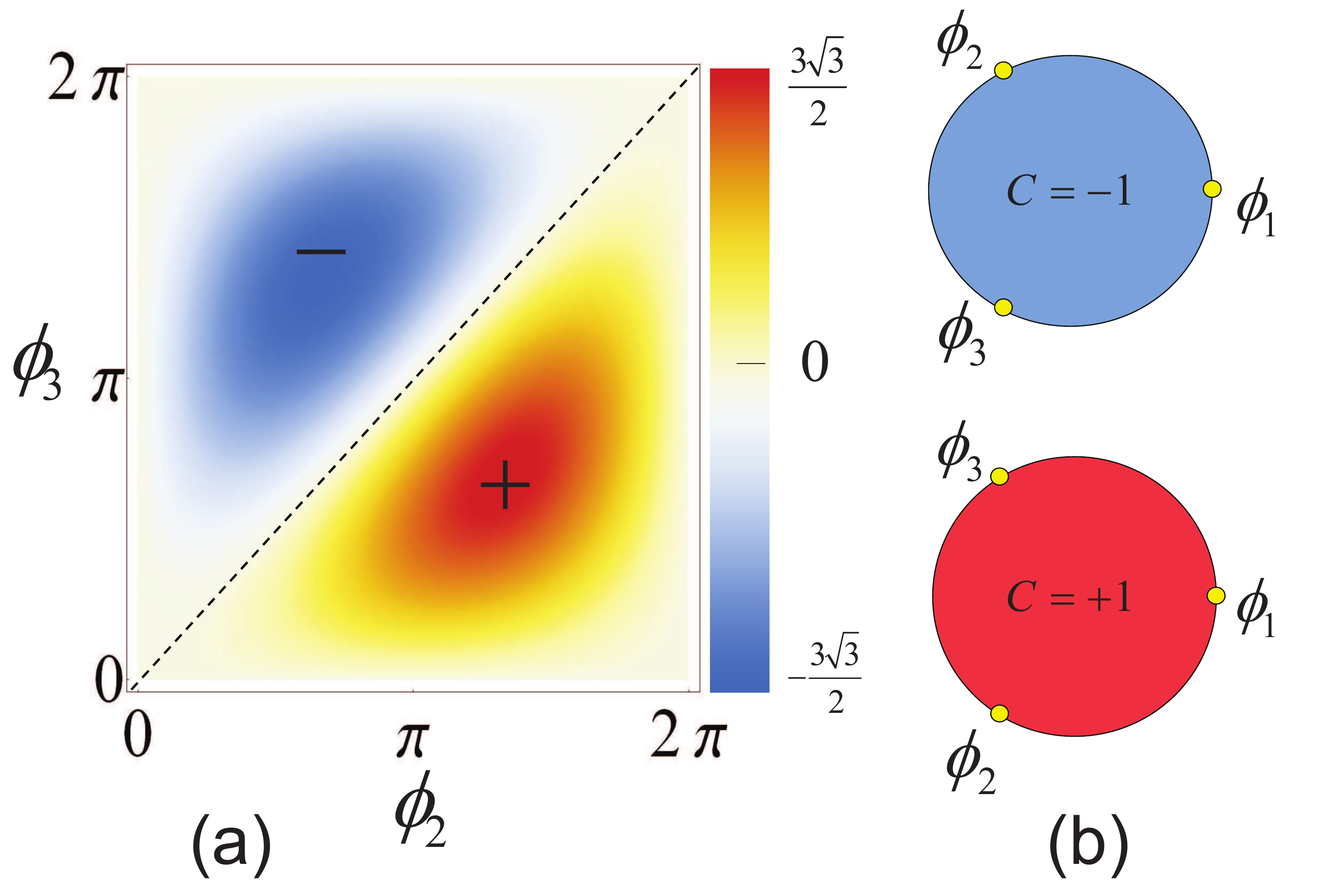, angle=0, width=0.48\textwidth}  
   \caption{{Topological phases of the superradiance lattice.} ({a}) The value of $\alpha$ (defined in Eq.(\ref{alpha})) as a function of the modulation phases of the coupling fields $\phi_2$ and $\phi_3$. $\phi_1=0$. The signs ``$\pm$" are for the Chern numbers $C=\pm1$. The dashed line separates the two topological phases. ({b}) The two Chern numbers corresponding to two topologically distinctive configurations of the three phase angles $\phi_1$, $\phi_2$ and $\phi_3$ on a unit circle. } 
  \label{flq} 
\end{figure}

We first show how to induce complex NNN hopping. We consider periodically modulated Rabi frequencies $\Omega_l$ $(l=1,2,3)$ of coupling fields,
\begin{equation}
\Omega_l=\Omega_s+2\Omega_d\cos(\nu_d t+\phi_l),
\label{cpl}
\end{equation}
where $\Omega_{s}$ and $\Omega_{d}$ are the static and dynamic components of the Rabi frequency, $\nu_d$ is the modulation frequency and $\phi_l$ is the modulation phase. Here we have assumed the size of the atomic ensemble be much smaller than $c/\nu_d$ (where $c$ is the speed of light) and hence neglected the position dependence of the modulation. On the other hand, the ensemble size is much larger than $c/\nu_c$ and the number of atoms $N\gg 1$ such that the TDS in the SL are approximately orthogonal to each other, i.e., $\la e_{\mathbf{k}^\prime}|e_{\mathbf{k}}\ra\approx \delta_{\mathbf{k}\mathbf{k}^\prime}$. The periodic modulation induces Floquet sidebands \cite{Shirley1965} with energy separation $\hbar\nu_d$. While $\Omega_s$ is the intra-sideband nearest-neighbour hopping, $\Omega_d$ induces the inter-sideband hopping. We choose $\nu_d\gg \Omega_{s/d},\Delta_c$ so that the Floquet sidebands are well separated in energy. Thus by second-order perturbation, the effective Hamiltonian of the NNN intra-sideband transition mediated by $\Omega_d$ is (see Supplementary Information)
\begin{equation}
\begin{aligned}
H^\prime=&\sum\limits_{\mathbf{k}}\sum\limits_{l\ne j=1}^{3}\hbar\Omega_{lj}
(|e_{\mathbf{k}+\mathbf{k}_l-\mathbf{k}_j}\ra \la e_{\mathbf{k}}|\\
&+|m_{\mathbf{k}-\mathbf{k}_1}\ra\la m_{\mathbf{k}-\mathbf{k}_1+\mathbf{k}_l-\mathbf{k}_j}|),
\end{aligned}
\end{equation}
where the NNN hopping coefficient
$\Omega_{lj}=2i\Omega^\prime\sin(\phi_l-\phi_j)$
with $\Omega^\prime=\Omega^2_d/\nu_d$. The fact that the NNN hopping coefficient is  purely imaginary is crucial to the topological phases. The loop transitions via NNN hopping accumulate nonzero phases, as shown in Fig.\ref{graphene} (c). $H^\prime$ opens a band gap $\epsilon_{tr}=4\sqrt{3}\hbar|\alpha|\Omega^\prime$ where $\alpha$ is a dimensionless quantity determined by the summation of $\Omega_{(l+1)l}$,
\begin{equation}
\alpha=-\sum\limits_{l=1}^3\sin(\phi_{l+1}-\phi_l)=4\prod\limits_{l=1}^3\sin\frac{\phi_{l+1}-\phi_l}{2}.
\label{alpha}
\end{equation}
The Chern numbers of the upper and lower bands $C$ and $C^\prime$ are $C=-C^\prime=\text{sign}(\alpha)$ (see Supplementary Information).
In Fig. \ref{flq} (a), we plot $\alpha$ with $\phi_1=0$ and $0\le\phi_{2,3}<2\pi$. The topological property of this SL Haldane model can be represented by the distribution of $\phi_l$ on a unit circle. There are two distinct topological configurations, counter-clockwise $\phi_1$, $\phi_2$ and $\phi_3$ for $C=-1$, and clockwise $\phi_1$, $\phi_2$ and $\phi_3$ for $C=1$, as shown in Fig.\ref{flq} (b). The time reversal $t\rightarrow-t$ in Eq.(\ref{cpl}) is equivalent to $\phi_l\rightarrow-\phi_l$, which leads to $C\rightarrow-C$.

\begin{figure} [t!]
\epsfig{figure=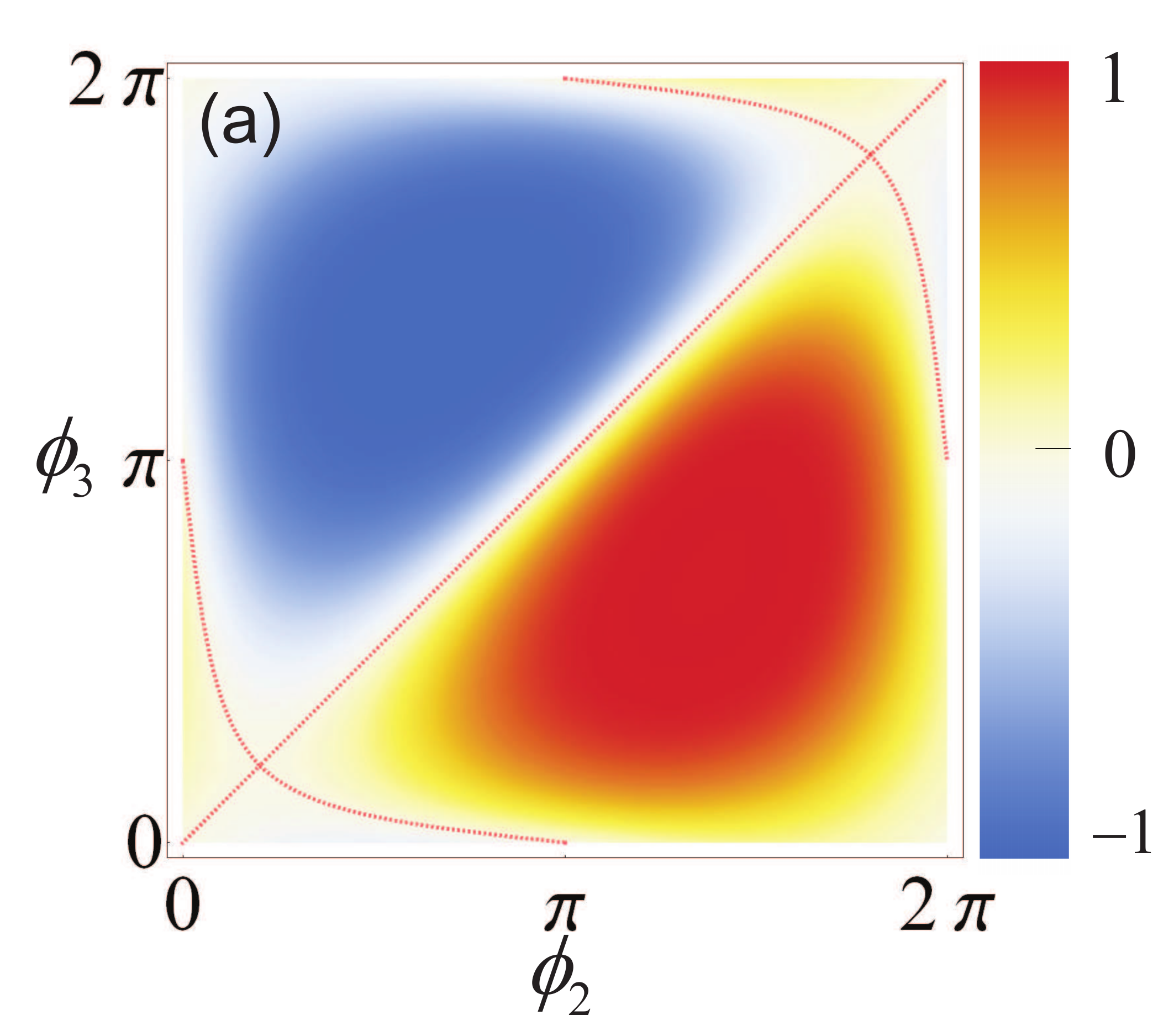, angle=0, width=0.4\textwidth}
\epsfig{figure=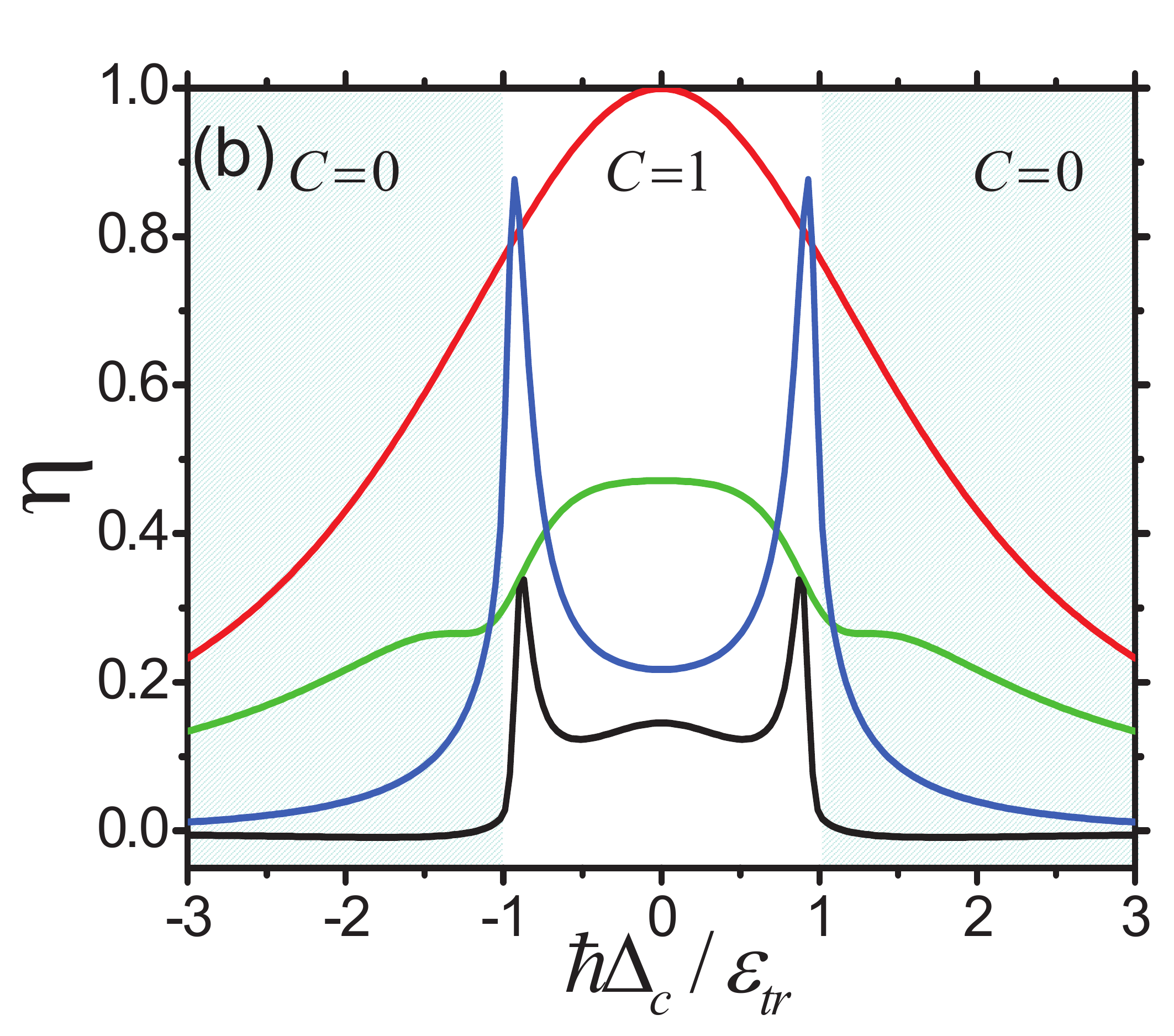, angle=0, width=0.4\textwidth}
\epsfig{figure=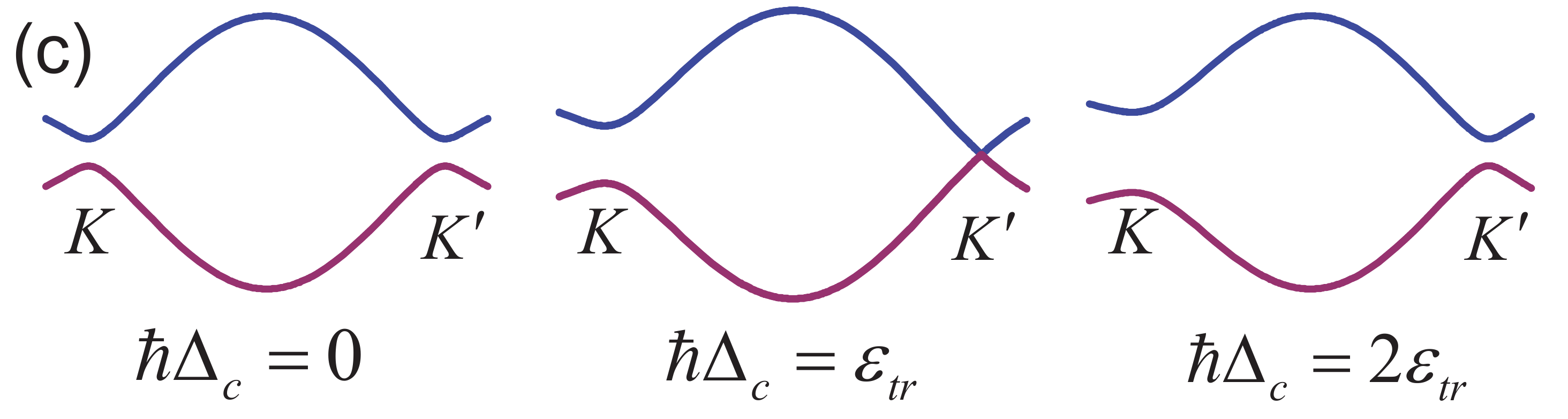, angle=0, width=0.45\textwidth}
\caption{{Topological phase transitions signatured by superradiance contrast.} ({a}) The contrast $\eta$ (defined in equation (6)) as a function of the modulation phases $\phi_2$ and $\phi_3$ with $\phi_1=0$. The red dot lines are zero points. $\Delta_p=\Delta_c=0$. $\Omega^\prime=0.01$. ({b}) The topological phase transition between $C=1$ and $0$. $\Delta_p=\Delta_c/2$. $\Omega^\prime=$ 0.01 (red), 0.1 (green), 0.5 (blue) and 1.0 (black). ({c}) The SL band structures in the topological phase transition. $\Omega^\prime=0.1$. $K$ and $K^\prime$ are the two valley points. For (b) and (c), $\phi_1=0$, $\phi_2=4\pi/3$ and $\phi_3=2\pi/3$. For all three figures, $\Omega_s=3$, $\gamma_e=1$ and $\gamma_m=0.1$.} 
\label{pd} 
\end{figure}

Unlike TIs \cite{Konig2007,Chang2013}, topological superradiance lattices (TSLs) have no outer edges in the semiclassical limit of the coupling fields (see Supplementary Information). Neither do TSLs have Fermi surfaces. Nonetheless, the TSL has its unique topological properties that are observable. 
The TDS have directional superradiance emission. Of all the TDS in the SL, only those $|e_\mathbf{k}\ra$ with $c|\mathbf{k}|-\omega_{eg}$ within the energy bands ($\omega_{eg}$ is the transition frequency between $|e\ra$ and $|g\ra$) can satisfy both energy and momentum conservation, and have directional emission in $\mathbf{k}$ \cite{Scully2006}. We call these states superradiant TDS and the other ones subradiant TDS. We can regard these superradiant TDS as an inner edge of a honeycomb lattice of subradiant TDS \cite{Lumer2013}. The topological orders lead to different light emissions from different superradiant TDS. Alternatively, we can also tune the probe field frequency to test the topological band properties at certain energy, which is analogue to tuning the Fermi surface in a fermionic system.

For the sake of simplicity, we set the wavevectors of the three EIT coupling fields to be $\mathbf{k}_1=-k_c\hat{x}$, $\mathbf{k}_2=k_c(\hat{x}-\sqrt{3}\hat{y})/2$ and $\mathbf{k}_3=k_c(\hat{x}+\sqrt{3}\hat{y})/2$, and the probe field wavevector $\mathbf{k}_p=-\mathbf{k}_1$. In this case we have only three superradiant TDS with wavevectors $\mathbf{k}_p$, $\mathbf{k}_+=-\mathbf{k}_3$ and $\mathbf{k}_-=-\mathbf{k}_2$, and the diffraction fields are along $\mathbf{k}_\pm$, as shown in Fig.\ref{graphene} (a). We set all fields on resonance, i.e., $\Delta_c=0$ and $\Delta_p\equiv\omega_{eg}-\nu_p=0$. The excitation $|e_{\mathbf{k}_p}\ra$ flows to $|e_{\mathbf{k}_\pm}\ra$ and emits photons along $\mathbf{k}_\pm$. We denote the steady state probability amplitudes of states $|e_{\mathbf{k}_{\pm}}\ra$ as $c_{\mathbf{k}_\pm}$ and define the superradiance contrast
\begin{equation}
\eta=\frac{|c_{\mathbf{k}_+}|^2-|c_{\mathbf{k}_-}|^2}{|c_{\mathbf{k}_+}|^2+|c_{\mathbf{k}_-}|^2}.
\end{equation}
For $C=1$, the excitation current flows along $|e_{\mathbf{k}_p}\ra \rightarrow|e_{\mathbf{k}_+}\ra \rightarrow |e_{\mathbf{k}_-}\ra$ \cite{Lumer2013}, as shown in Fig.\ref{graphene} (c). Since each TDS $|e_\mathbf{k}\ra$ or $|m_\mathbf{k}\ra$ has decoherence rate $\gamma_e$ or $\gamma_m$, respectively, the excitation decays while flowing and it is more probable in state $|e_{\mathbf{k}_+}\ra$ than in state $|e_{\mathbf{k}_-}\ra$. We therefore have $\eta>0$. Similarly, for $C=-1$, $\eta<0$. Thus the sign change of the superradiance contrast signatures the topological phase transition, as seen in Fig.\ref{pd} (a). The superradiance contrast in Fig.\ref{pd} (a) is consistent with Fig.\ref{flq} (b) except for the two diagonal corners where the topological currents are weak and the local effect inside a unit cell dominates (see Supplementary Information).

\begin{figure}
\epsfig{figure=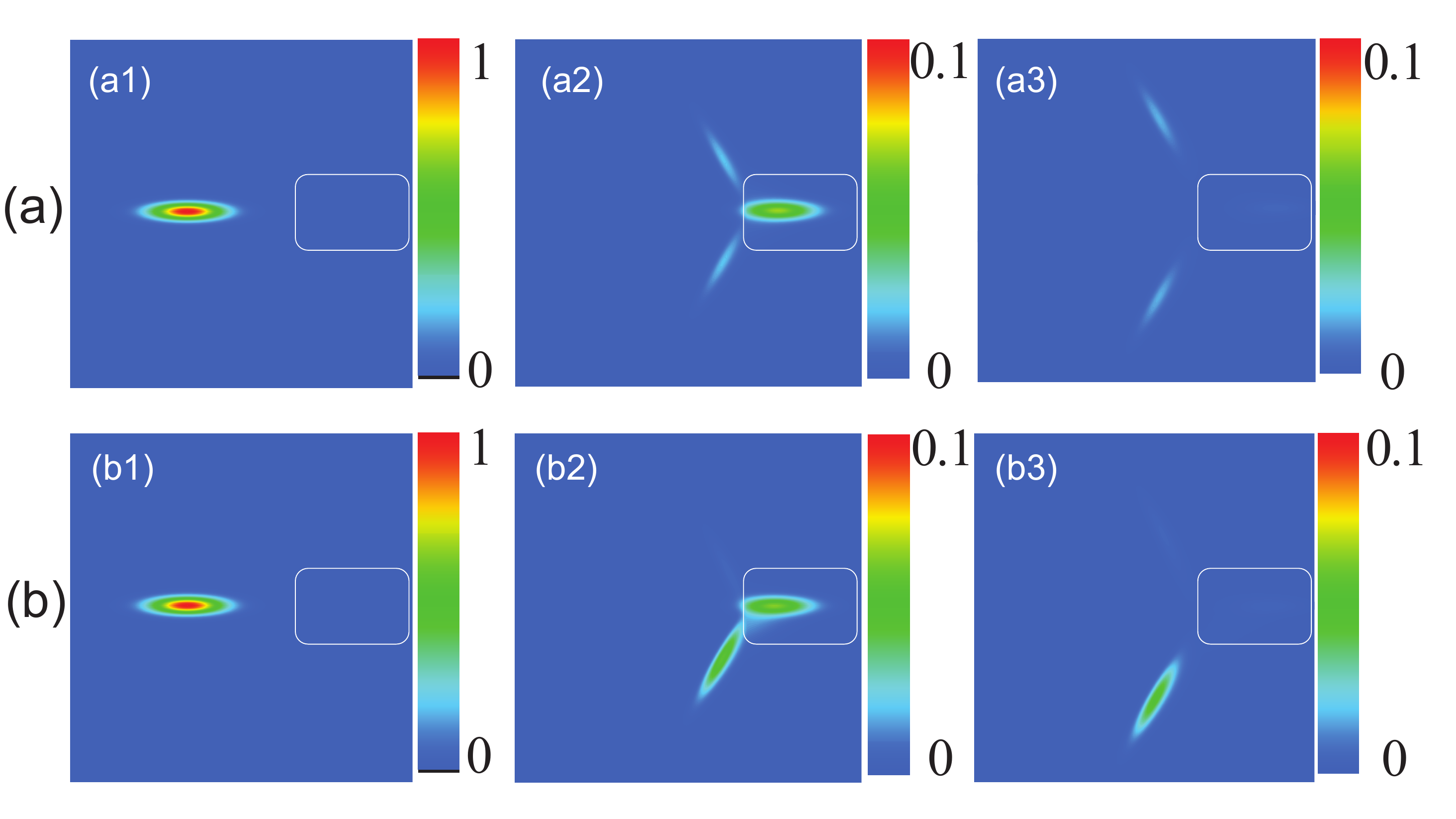, angle=0, width=0.5\textwidth}
\caption{{Real-space propagation of a light pulse through an atom ensemble driven into a trivial or topological superradiance lattices.} Diffraction of a weak Gaussian probe pulse in x-y plane for ({a}) a trivial SL with $\Omega_d=0$ and ({b}) a TSL with $C=1$. $\phi_1=0$, $\phi_2=4\pi/3$, $\phi_3=2\pi/3$ and $\Omega^\prime=0.01$. The derived linear and nonlinear susceptibilities are $\chi^{(1)}=i0.1410$, $\chi_+=\chi_-=-i0.0522$ in (a) and $\chi^{(1)}=i0.1446$, $\chi_+=-i0.1057$, $\chi_-=-i0.0021$ in (b). The parameters are such that $\Omega_s= 3$, $\Delta_p=\Delta_c=0$, $\gamma_e=1$ and $\gamma_m=0.1$. The square denotes the area occupied by atoms. The three figures in each group are for scaled time $t=1,80,100$ successively. We assume the group velocity be uniform everywhere.} 
\label{pulse} 
\end{figure}
 
The topological phase with $C=0$ for non-zero $\Omega^\prime$ can be reached by breaking the inversion symmetry. Substantially easier than in graphene electrons, a sublattice offset in SLs can be introduced by choosing $\Delta_c\ne 0$. We set $\phi_l=4\pi(l-1)/3$ and thus $C=1$ when $\Delta_c=0$. The energy gap in absence of sublattice offset is $\epsilon_{tr}=18\hbar\Omega^\prime$. Topological phase transitions occur at $\hbar|\Delta_c|=\epsilon_{tr}$ \cite{Haldane1988} (see Supplementary Information). In Fig.\ref{pd} (b), we plot $\eta$ as a function of $\hbar\Delta_c/\epsilon_{tr}$. For $\Omega^\prime=0.01$, the phase transition is smeared out by the relatively large $\gamma_e$. The phase transition gets more apparent for larger $\Omega^\prime$. Depending on $\Omega^\prime$, $\eta$ has peaks or kinks near the phase transition point because the two bands touch at the middle of the band gap, where the probe field probes, as shown in Fig.\ref{pd} (c).

The TSLs have unique features in transient light propagation under pulse probe. In Fig.\ref{pulse}, we compare the pulse propagation in a trivial SL with zero $\Omega^\prime$ and in a TSL. For a weak probe pulse, the linear susceptibility is $\chi^{(1)}\propto c_{\mathbf{k}_p}$ and the linear absorption is $\text{Im}\chi^{(1)}$. The two multi-wave-mixing signals along $\mathbf{k}_\pm$ correspond to the nonlinear susceptibilities $\chi_\pm\propto c_{\mathbf{k}_\pm}$ and can be understood as a result of optical grating \cite{Wang2015}.  We simulate the pulse propagation for the three modes along $\mathbf{k}_{p,\pm}$ using coupled wave equations \cite{Wang2013} (see Supplementary Information). For a trivial SL without modulation, the light propagating along $\mathbf{k}_\pm$ is symmetric, while for a TSL with $C=1$, the topological currents drive the probe pulse to $\mathbf{k}_+$, even if the NNN hopping is two orders of magnitude smaller than the nearest-neighbour hopping.

TSLs can be readily realised in experiments for cold alkali atoms. Taking $^{85}$Rb D1 line for example, we can have $|g\ra=|5^{2}S_{1/2}, F=2\ra$, $|e\ra=|5^{2}P_{1/2}, F=2\ra$ and $|m\ra=|5^{2}S_{1/2}, F=3\ra$. $\gamma_e=2.9$MHz and $\gamma_m$ is controllable via inhomogeneous magnetic field \cite{Tiwari2010}. The Rabi frequency $\Omega_s=3\gamma_e=8.6$MHz (intensity 25mW/cm$^{2}$). The modulation frequency can be $\nu_d=10\gamma_e=28.8$MHz which is large enough to separate the Floquet bands. One can trap $10^6$ atoms in 1 mm$^3$ such that $N\gg 1$ and the size $L$ of the ensemble $c/\nu_c\ll L\ll c/\nu_d$. In the $\mu$K regime, the thermal random motions have negligible Doppler shifts ($\sim$kHz). Another possible type of physical systems are rare earth atoms doped in solids \cite{Thiel2011}. One should first optically pump nearly all population to $|g\ra$, then turn on three optical fields coupling $|e\ra$ to $|m\ra$, send in a weak field probing the $|g\ra$ to $|e\ra$ transition, and detect the diffraction signals.

Similar to the electronic TIs \cite{Lindner2011, Kitagawa2010} and the optical lattice simulations of TIs \cite{Jotzu2014}, the topological properties of TSLs are determined by Schr\"odinger equations, different from the photonic TIs \cite{Haldane2008,HafeziM2013,Rechtsman2013,Khanikaev2013}, which are governed by Maxwell equations. The generation and detection of the topological properties of TSLs, however, can be easily controlled by light. The unique feature of the TSL is that its lattice sites have discrete momenta rather than positions. It has the advantage to be extended to dimensions higher than three where no real space lattices exist \cite{Wang2015} and offers a platform for high-dimensional topological physics \cite{Zhang2001}.

\section*{Funding Information}
\noindent Science Foundation Grants No. PHY-1241032 (INSPIRE CREATIV) and PHY-1068554; Robert A. Welch Foundation (Grant No. A-1261); Herman F. Heep and Minnie Belle Heep Texas A\&M University Endowed Fund; Hong Kong RGC/GRF; CUHK VC's One-Off Discretionary Fund; Natural Science Foundation of China U1330203.\\

The authors thank Marlan O. Scully for helpful discussions.

\bibliographystyle{apsrev4-1}

\begin{center}
\textbf{\Large{Supplementary I}}
\end{center}

\setcounter{equation}{0}
\renewcommand{\theequation}{S.\arabic{equation}}
\setcounter{figure}{0}
\renewcommand{\thefigure}{S.\arabic{figure}}

\section{Effective Hamiltonian}
\label{sc1}

The three-level atoms we use to construct the two-dimensional (2D) superradiance lattice (SL) have a ground state $|g\ra$, an excited state $|e\ra$ and a third state $|m\ra$. The optical fields that couple $|e\ra$ and $|m\ra$ have three modes with wave vectors $\mathbf{k}_1=-k_c\hat{x}$, $\mathbf{k}_2=k_c(\frac{1}{2}\hat{x}-\frac{\sqrt{3}}{2}\hat{y})$ and $\mathbf{k}_3=k_c(\frac{1}{2}\hat{x}+\frac{\sqrt{3}}{2}\hat{y})$. The interaction Hamiltonian with rotating-wave approximation is
\begin{equation}
H_c=-\hbar\sum\limits_{l=1}^{3}\sum\limits_{j=1}^{N}\kappa_l a_l e^{i\mathbf{k}_l\cdot \mathbf{r}_j}|e_j\ra\la m_j|+H.c.,
\label{Hc}
\end{equation}
where $\kappa_l$ and $a_l$ are the vacuum coupling strength and annihilation operator of the $l$th mode, respectively. $N$ is the number of atoms and $\mathbf{r}_j$ is the position of the $j$th atom.

\begin{figure} 
\epsfig{figure=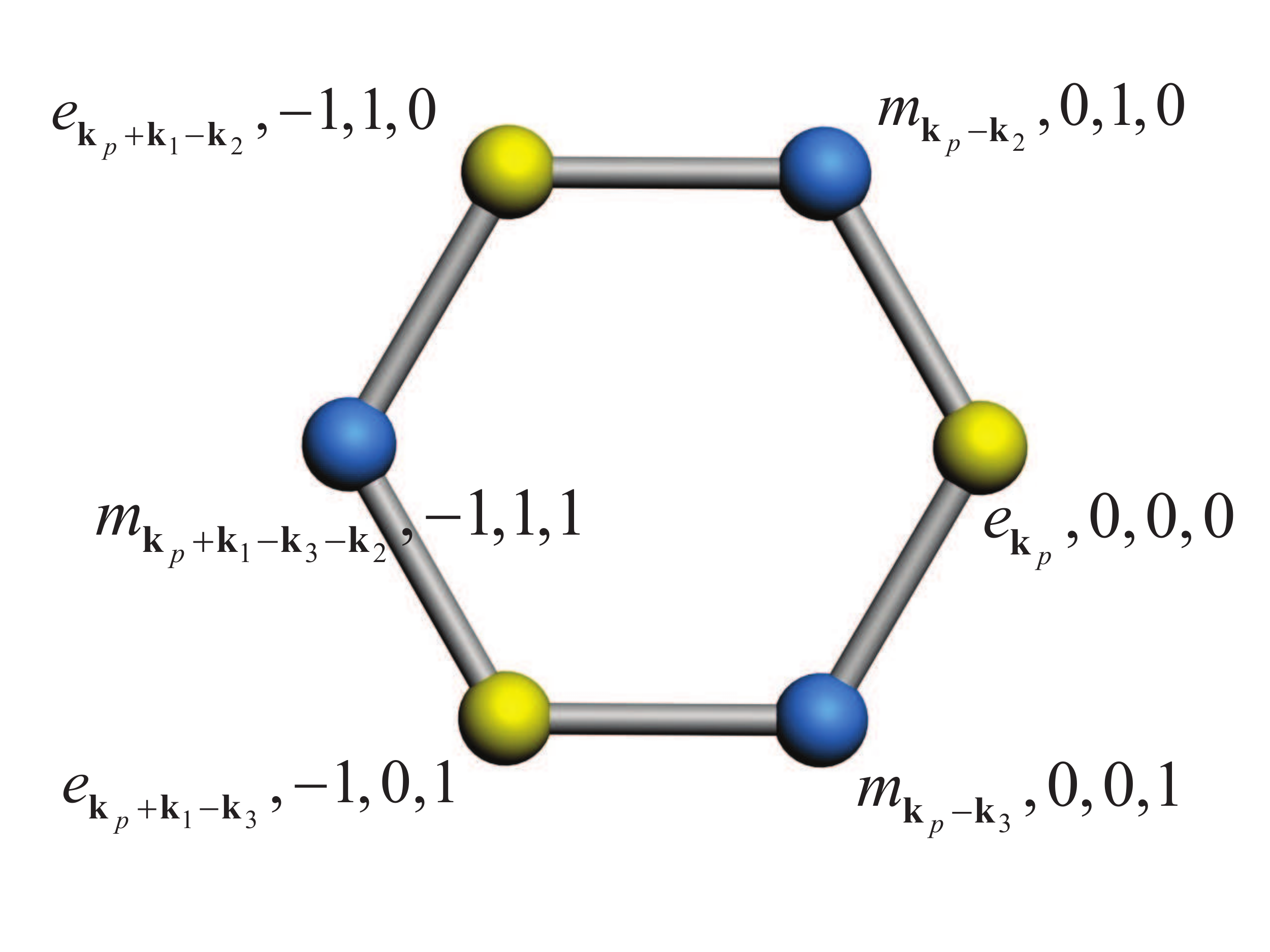, angle=0, width=0.45\textwidth}   
  \caption{{Timed Dicke states and the associated photon numbers in an SL.} The numbers associated with each TDS are the photon number difference from those of state $|e_{\mathbf{k}_p}\ra$.} 
  \label{s1} 
\end{figure}

Initially the atomic ensemble is in the ground state $|G\ra=|g_1, g_2,...,g_N\ra$. A weak probe field with wave vector $\mathbf{k}_p$ can prepare the atomic ensemble in the timed Dicke state (TDS)
\begin{equation}
|e_{\mathbf{k}_p}\ra=\frac{1}{\sqrt{N}}\sum\limits_{j=1}^{N}e^{i\mathbf{k}_p\cdot \mathbf{r}_j}|g_1, g_2,...,e_j,...,g_N\ra
\label{ek}
\end{equation} 
by the collective absorption of a single photon. The combinational quantum states of the atoms and three coupling modes can be written as $|e_{\mathbf{k}_p},n_1,n_2,n_3\ra$, where $n_l$ $(l=1,2,3)$ is the photon number in mode $\mathbf{k}_l$. This state is coupled to $|m_{\mathbf{k}_p-\mathbf{k}_2},n_1,n_2+1,n_3\ra$ with coupling strength $-\hbar\kappa_2\sqrt{n_2+1}$. The state $|m_{{\mathbf k}_p-{\mathbf k}_2}\ra$ is defined the same as in equation (\ref{ek}) with $e_j$ replaced by $m_j$ and ${\mathbf k}_p$ replaced by ${\mathbf k}_p-{\mathbf k}_2$. Similar coupling also exists for the other two modes. $|m_{\mathbf{k}_p-\mathbf{k}_2},n_1,n_2+1,n_3\ra$ is in turn coupled to $|e_{\mathbf{k}_p+\mathbf{k}_1-\mathbf{k}_2},n_1-1,n_2+1,n_3\ra$ or $|e_{\mathbf{k}_p+\mathbf{k}_3-\mathbf{k}_2},n_1,n_2+1,n_3-1\ra$ via excitation by modes $\mathbf{k}_1$ or $\mathbf{k}_3$. These states form a honeycomb lattice with discrete momentum coordinates, as shown in Fig. \ref{s1}, called the superradiance lattice \cite{2014arXiv1403.7097W}. The two sublattices of the SL correspond to TDS for $|e\ra$ and $|m\ra$. We denote the coupling field frequency as $\nu_c$ and the transition frequency between $|e\ra$ and $|m\ra$ as $\omega_{em}$. The energy difference between the two sublattice is $\Delta_c=\omega_{em}-\nu_c$. We set the zero energy at the middle of the energies of the two sublattices. Then the energies of the $|e\ra$ and $|m\ra$ sublattices are $\Delta_c/2$ and $-\Delta_c/2$, respectively. 

Although the coupling strengths in the SL are site-dependent, if we use coherent coupling fields with large average photon numbers $\left\langle n_l\right\rangle\gg 1$, the Rabi frequency of the $l$th mode can be approximated as the classical Rabi frequency $\Omega_l=\kappa_l\sqrt{\left\langle n_l\right\rangle}$. The SL with quantized photon numbers has edges when either of the three coupling field photon numbers reduces to zero. The total lattice has a triangular boundary. However, since the TDS have finite life time, which we suppose to be $\tau$ on average, once we create an excitation $|e_{\mathbf{k}_p}\ra$ in the SL, the average distance this excitation can travel is in the order of $\Omega_l\tau$. In this paper we have $\Omega_l\tau\ll \la n_l\ra$ and the edges can never be reached. Interesting edge effects might exist in the few-photon limit and it will be discussed elsewhere. Since the photon numbers are correlated to the momenta $\mathbf{k}$'s in the TDS, we can simplify the notations by dropping the photon numbers. The Hamiltonian of these TDS is
\begin{equation}
\begin{aligned}
H=&\frac{\hbar\Delta_c}{2}\sum\limits_{\mathbf{k}}(|e_\mathbf{k}\ra\la e_\mathbf{k}|-|m_{\mathbf{k}-\mathbf{k}_1}\ra\la m_{\mathbf{k}-\mathbf{k}_1}|)\\
&-\sum\limits_{\mathbf{k}}\sum\limits_{l=1}^{3}\hbar\Omega_l [|e_{\mathbf{k}}\ra\la m_{\mathbf{k}-\mathbf{k}_l}|+h.c.],
\end{aligned}
\label{Hh}
\end{equation}
where $\mathbf{k}=\mathbf{k}_p+r(\mathbf{k}_2-\mathbf{k}_1)+s(\mathbf{k}_3-\mathbf{k}_2)$ with integers $r$ and $s$. 

The next-nearest-neighbour (NNN) hopping terms are induced by the periodic modulation of the three coupling field Rabi frequencies,
\begin{equation}
\Omega_l=\Omega_s+2\Omega_d\cos(\nu_d t-\mathbf{k}_d^l\cdot\mathbf{r}+\phi_l),
\label{cpl}
\end{equation}
where $\Omega_{s}$ and $\Omega_{d}$ are the static and dynamic components of the Rabi frequencies. $\nu_d$ is the modulation frequency. $\phi_l$ is the modulation phase of the $l$th field. $\mathbf{k}_d^l$ is the modulation wavevector. We choose the atomic ensemble much smaller than $1/|\mathbf{k}_d^l|$ such that the position dependence of the phase $-\mathbf{k}_d^l\cdot\mathbf{r}$ can be neglected. We set $-\mathbf{k}_d^l\cdot\mathbf{r}=0$ for simplicity. On the other hand, the size of the atomic ensemble shall be much larger than $1/\nu_c$ and the total number of atoms $N\gg 1$ such that
\begin{equation}
\la e_{\mathbf{k}^\prime}|e_\mathbf{k}\ra=\frac{1}{N}\sum\limits_{j=1}^{N}e^{i(\mathbf{k}-\mathbf{k}^\prime)\cdot\mathbf{r}_j}\approx \delta_{\mathbf{k}\mathbf{k}^\prime}.
\end{equation}
We assume the atoms be randomly distributed and their number is large enough to cover all possible points in the real space Brillouin Zone. The SL can be regarded as infinite. 

We expand the Hamiltonian into static, positive- and negative-frequency components, 
\begin{equation}
H=H_0+H_{+1}e^{i\nu_dt}+H_{-1}e^{-i\nu_dt},
\label{hf}
\end{equation}
where
\begin{equation}
\begin{aligned}
H_0=&\frac{\hbar\Delta_c}{2}\sum\limits_{\mathbf{k}}(|e_\mathbf{k}\ra\la e_\mathbf{k}|-|m_{\mathbf{k}-\mathbf{k}_1}\ra\la m_{\mathbf{k}-\mathbf{k}_1}|)\\
&-\hbar\Omega_s\sum\limits_{\mathbf{k}}\sum\limits_{l=1}^{3} [|e_{\mathbf{k}}\ra\la m_{\mathbf{k}-\mathbf{k}_l}|+h.c.],
\end{aligned}
\end{equation}
\begin{equation}
H_{+1}=-\hbar\Omega_d\sum\limits_{\mathbf{k}}\sum\limits_{l=1}^{3}e^{i\phi_l} [|e_{\mathbf{k}}\ra\la m_{\mathbf{k}-\mathbf{k}_l}|+h.c.],
\end{equation}
\begin{equation}
H_{-1}=-\hbar\Omega_d\sum\limits_{\mathbf{k}}\sum\limits_{l=1}^{3}e^{-i\phi_l} [|e_{\mathbf{k}}\ra\la m_{\mathbf{k}-\mathbf{k}_l}|+h.c.].
\end{equation}
Note that $H_{\pm 1}$ are not Hermitian themselves, but $H_{+1}$ is the Hermitian conjugate of $H_{-1}$. The phase factors $e^{\pm i\phi_l}$ in $H_{\pm 1}$ play the crucial role for the complex NNN hopping terms in the Haldane model. 

The dynamics of the system is a Floquet problem \cite{Shirley1965, Kitagawa2011}. Based on the Floquet theorem, the wave function can be written as
\begin{equation}
|\Psi\ra=e^{-i\epsilon t/\hbar}|\psi(t)\ra,
\end{equation}
where $\epsilon$ is the quasi-eigenenergy and 
\begin{equation}
|\psi(t)\ra=\sum_n e^{in\nu_dt}|\psi_n\ra,
\label{psit}
\end{equation}
is a periodic wave function with period $2\pi/\nu_d$. Substituting equations (\ref{hf})-(\ref{psit}) to the Schr\"odinger equation, $i\hbar\partial|\Psi\ra/\partial t=H|\Psi\ra$, we obtain
\begin{equation}
\begin{aligned}
&i\hbar\frac{\partial}{\partial t}e^{-i\epsilon t/\hbar}\sum_n e^{in\nu_dt}|\psi_n\ra\\
&=(H_0+H_{+1}e^{i\nu_dt}+H_{-1}e^{-i\nu_dt})e^{-i\epsilon t/\hbar}\sum_n e^{in\nu_dt}|\psi_n\ra.
\end{aligned}
\end{equation}
The terms with the same time evolution phase factors should be equal on both sides. We therefore have
\begin{equation}
(\epsilon-\hbar n\nu_d)|\psi_n\ra=H_0|\psi_n\ra+H_{+1}|\psi_{n-1}\ra+H_{-1}|\psi_{n+1}\ra.
\label{eflq}
\end{equation}
The quasi-eigenenergy can be obtained by diagonalizing the above Hamiltonian. 

For the sake of simplicity, we assume the separation between the Floquet sidebands is much larger than the bandwidth, $\nu_d\gg\Omega_{s,d},\Delta_c$, where the perturbation theory can be applied \cite{Liu2000,Gomez-Leon2013}. When the probe field is near resonance, $\Delta_p=\omega_{eg}-\nu_p\ll \Omega_s$ where $\omega_{eg}$ is the transition frequency between $|e\ra$ and $|g\ra$ and $\nu_p$ is the probe field frequency, only the Floquet band with $n=0$ in Eq.(\ref{eflq}) is relevant.
For states with eigenfrequencies near $\epsilon=0$, the effective Hamiltonian can be obtained  by standard second-order perturbation as 
\begin{equation}
\epsilon|\psi_0\ra=H_\text{eff}|\psi_0\ra,
\label{heff}
\end{equation}
where
\begin{equation}
H_{\text{eff}}=H_0+H^\prime,
\end{equation}
with NNN Hamiltonian
\begin{equation}
H^\prime=\frac{1}{\hbar\nu_d}(H_{+1}H_{-1}-H_{-1}H_{+1}).
\label{he}
\end{equation}
Explicitly,
\begin{equation}
\begin{aligned}
H^\prime=&\sum\limits_{\mathbf{k}}\sum\limits_{l\ne j=1}^{3}\hbar\Omega_{lj}\\
&\left(|e_{\mathbf{k}+\mathbf{k}_l-\mathbf{k}_j}\ra \la e_{\mathbf{k}}|+|m_{\mathbf{k}-\mathbf{k}_1}\ra\la m_{\mathbf{k}-\mathbf{k}_1+\mathbf{k}_l-\mathbf{k}_j}|\right),
\end{aligned}
\end{equation}
where $\Omega_{lj}=\Omega^\prime[e^{i(\phi_l-\phi_j)}-e^{i(\phi_j-\phi_l)}]=2i\Omega^\prime\sin(\phi_l-\phi_j)$ with $\Omega^\prime=\Omega^2_d/\nu_d$. The crucial factor $i =\sqrt{-1}$ comes from the quantum interference between the two pathways shown in Fig.\ref{flq} (a). The loop transitions via NNN hopping accumulate nonzero phases due to these complex factors, as shown in Fig.\ref{flq} (b).
\begin{figure} 
\epsfig{figure=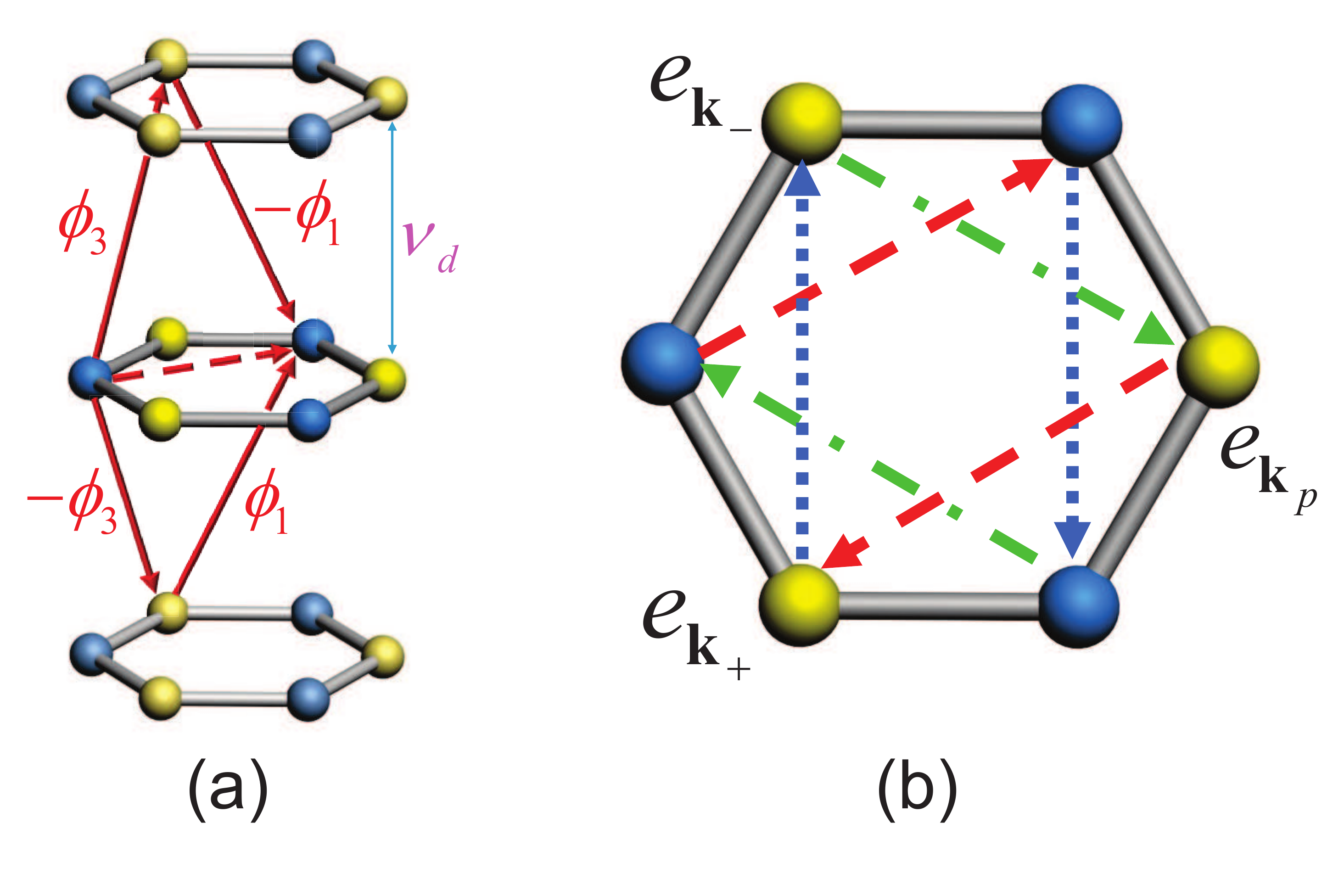, angle=0, width=0.5\textwidth}  
  \caption{{ Complex next-nearest-neighbor hopping induced by Rabi frequency modulation.} (a) The Floquet sidebands of the SL and the NNN transition $\Omega_{13}$ (dashed arrow) induced by the interference between two second-order inter-sideband transitions (solid arrow). Phases are labelled for each step. (b) The effective NNN transitions in a unit cell. The NNN hopping $\Omega_{31}$ (red arrow), $\Omega_{23}$ (blue dot arrow) and $\Omega_{12}$ (green dash dot arrow) enclose a nonzero effective magnetic flux.} 
  \label{flq} 
\end{figure}

The effective Hamiltonian is greatly simplified in the real-space representation. We denote the real-space basis states as
\begin{equation}
\begin{aligned}
|e_{\mathbf{r}_j}\rangle&=|g_1,g_2,\ldots, e_j,\ldots, g_N\rangle,\\
|m_{\mathbf{r}_j}\rangle&=|g_1,g_2,\ldots, m_j,\ldots, g_N\rangle,
\end{aligned}
\label{ft}
\end{equation}
The effective Hamiltonian can be written as
\begin{equation}
H_{\text{eff}}=\sum_{j}{\mathbf h}({\mathbf r}_j)\cdot{\boldsymbol\sigma}_j 
\label{hr}
\end{equation}
where the effective magnetic field ${\mathbf h}({\mathbf r}_j)=(h_x,h_y,h_z)$ with 
\begin{equation}
h_x=-\hbar\Omega_s\sum\limits_{l=1}^3\cos(\mathbf{r}_j\cdot\mathbf{k}_l),
\label{hx}
\end{equation}
\begin{equation}
h_y=\hbar\Omega_s\sum\limits_{l=1}^3\sin(\mathbf{r}_j\cdot\mathbf{k}_l),
\label{hy}
\end{equation}
\begin{equation}
h_z=\frac{\hbar\Delta_c}{2}+2i\hbar\sum\limits_{l=1}^3\Omega_{(l+1)l}\sin[\mathbf{r}_j\cdot(\mathbf{k}_{l+1}-\mathbf{k}_{l})],
\label{hz}
\end{equation}
and the pseudo spin ${\boldsymbol\sigma}_j=(\sigma_j^x,\sigma_j^y,\sigma_j^z)$ with the Pauli matrices for the $j$th atom defined as $\sigma_j^x=|e_{{\mathbf r}_j}\rangle\langle m_{{\mathbf r}_j}|+|m_{{\mathbf r}_j}\rangle\langle e_{{\mathbf r}_j}|$, $\sigma_j^y=-i|e_{{\mathbf r}_j}\rangle\langle m_{{\mathbf r}_j}|+i|m_{{\mathbf r}_j}\rangle\langle e_{{\mathbf r}_j}|$ and $\sigma_j^z=|e_{{\mathbf r}_j}\rangle\langle e_{{\mathbf r}_j}|-|m_{{\mathbf r}_j}\rangle\langle m_{{\mathbf r}_j}|$.

\section{Berry connection and Berry curvature}
\label{sc3}

The topological properties of the wavefunctions are determined by the Berry connection $\mathbf{A}$ and Berry curvature $\mathbf{B}$,
\begin{equation}
\mathbf{A}=i\la\psi|\nabla|\psi\ra,
\end{equation}
\begin{equation}
\mathbf{B}=\nabla\times\mathbf{A},
\end{equation}
where $\nabla$ and $\nabla\times$ are the gradient and curl operators with respect to $\mathbf{r}$.

\begin{figure} 
\epsfig{figure=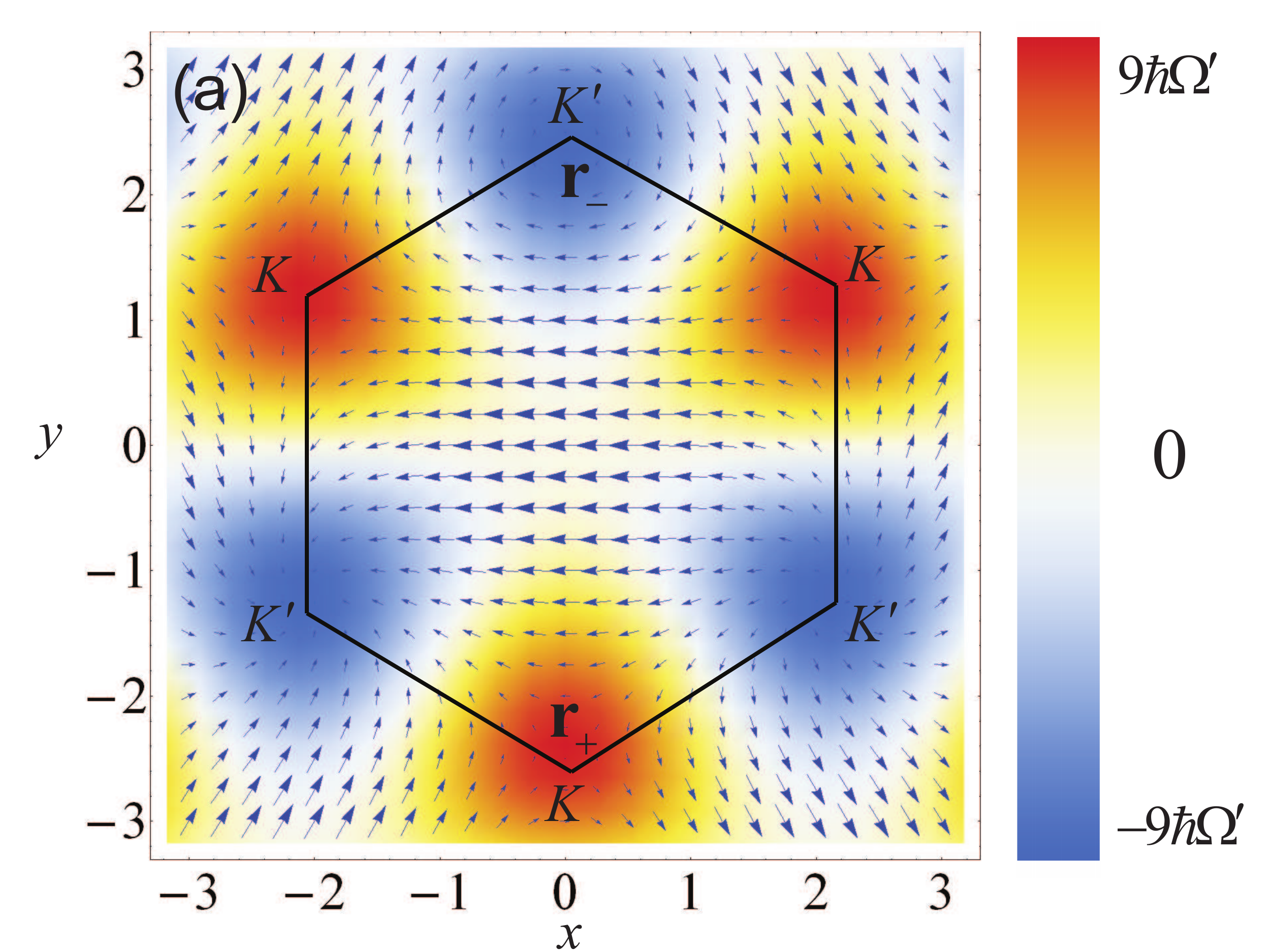, angle=0, width=0.45\textwidth}  
\epsfig{figure=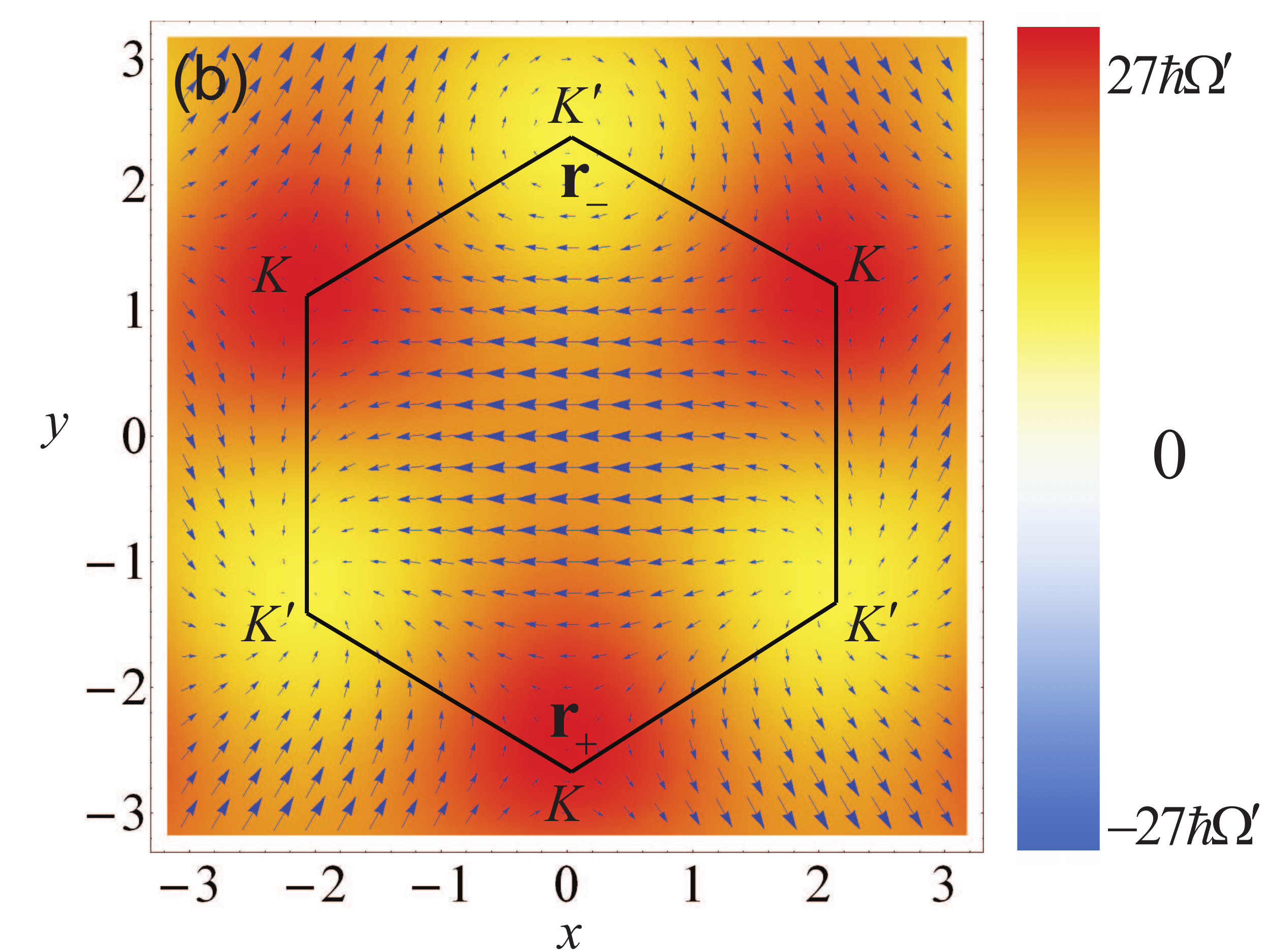, angle=0, width=0.45\textwidth} 
  \caption{{ Effective magnetic fields in a Brillouin zone for topological and trivial SL's.} (a) A topological SL with $\Delta_c=0$. (b) A trivial SL with $\Delta_c=36\Omega^\prime$. The arrows show the vector $h_x\hat{x}+h_y\hat{y}$ and the colours show $h_z$. The $x$ and $y$ axis are in unit of $k^{-1}_c$. The phase of the coupling field $\mathbf{k}_l$ is $\phi_l=(l-1)4\pi/3$ and $\alpha=3\sqrt{3}/2$. Note that the $K/K^\prime$ points at the boundaries of the first Brillouin zone are equivalent.} 
  \label{effb} 
\end{figure}

Similar to a spin-1/2 in a magnetic field ${\mathbf h}$, the eigenenergies of the SL eigenstates of the effective Hamiltonian in equation (\ref{hr}) are 
\begin{equation}
\epsilon_\pm=\pm h
\end{equation}
where $\pm$ are for the upper and lower bands and $h$ is the magnitude of $\mathbf{h}$.
The eigen wavefunction in the upper band can be written as
\begin{equation}
|\psi_+(\mathbf{r})\ra=\frac{1}{\sqrt{2h(h+h_z)}}
\left(
\begin{array}{ccc}
h+h_z\\
h_x+ih_y
\end{array}
\right).
\label{psi1}
\end{equation}
This wavefunction is well defined except for the south pole of the Bloch sphere where $\mathbf{h}=(0,0,h_z)$ with $h_z=-h$. The eigen wavefunction near the south pole can be written as 
\begin{equation}
|\psi_+^\prime(\mathbf{r})\ra=e^{i\phi(\mathbf{r})}|\psi_+(\mathbf{r})\ra=\frac{1}{\sqrt{2h(h-h_z)}}
\left(
\begin{array}{ccc} 
h_x-ih_y \\ 
h-h_z
\end{array}
\right),
\end{equation}
where the gauge transformation
\begin{equation}
e^{i\phi({\mathbf{r}})}=\frac{h_x-ih_y}{|h_x-ih_y|}.
\end{equation}
This way, the whole Bloch sphere is fully covered by two gauges \cite{Wu1975}.

The Berry connection in the upper band (except for the south pole of the Bloch sphere) is
\begin{equation}
\begin{aligned}
\mathbf{A}_+=&i\la\psi_+|\nabla|\psi_+\ra\\
=&-\frac{1}{2h(h+h_z)}\left(h_x\nabla h_y-h_y\nabla h_x\right),
\end{aligned}
\end{equation}
and near the south pole we can use
\begin{align}
\mathbf{A}^\prime_+=\mathbf{A}_+-\nabla\phi(\mathbf{r}).
\end{align}
The Berry curvature is
\begin{equation}
\begin{aligned}
\mathbf{B}_+=&\nabla\times\mathbf{A}_+\\
=&-2\hat{z}\text{Im}\la\partial_x\psi_+|\partial_y\psi_+\ra\\
=&-\frac{\hat{z}}{2h^3}\epsilon_{abc}h_a\partial_xh_b\partial_yh_c,
\end{aligned}
\end{equation}
where $\epsilon_{abc}$ with $a,b,c\in{x,y,z}$ is the Levi-Civita symbol.
The Chern number is defined as the total Berry curvature in the whole first Brillouin zone
\begin{equation}
\begin{aligned}
C=&\frac{1}{2\pi}\oiint_{\text{BZ}} \mathbf{B}_+\cdot\,d\mathbf{S}\\
=&-\frac{1}{4\pi}\oiint_{\text{BZ}} \frac{1}{h^3}\epsilon_{abc}h_a\partial_xh_b\partial_yh_c \,dx \,dy,
\end{aligned}
\end{equation}
which counts the winding number of the effective magnetic field $\mathbf{h}$ wrapping around the Bloch sphere in the whole Brillouin zone \cite{Qi2006,Katan2013}.

In Fig.\ref{effb}, we plot $\mathbf{h}$ in the Brillouin zone for a topological nontrivial SL where the band gap is opened by the NNN hopping. If we go from the $K$ point $\mathbf{r}_+=-\frac{4\sqrt{3}\pi}{9k_c}\hat{y}$ to the $K^\prime$ point $\mathbf{r}_-=\frac{4\sqrt{3}\pi}{9k_c}\hat{y}$, $\mathbf{h}$ moves from the north pole to the south pole of the Bloch sphere. The Chern number is one. If the band gap is opened by the on-site offset, $\Delta_c$, $h_z$ is a constant in the Brillouin zone. $\mathbf{h}$ can only cover a patch on the Bloch sphere and the Chern number is zero.

The two bands of $H_{\text{eff}}$ have the smallest gap at the $K$ and $K^\prime$ points $\mathbf{r}_\pm$, where
\begin{equation}
\sum\limits_{l=1}^{3} e^{i\mathbf{r}_\pm\cdot\mathbf{k}_l}=0,
\end{equation} 
and hence $h_x=h_y=0$. At these symmetry points of the Brillouin zone the Hamiltonian is diagonal,
\begin{equation}
\begin{aligned}
\mathcal{H}(\mathbf{r}_\pm)=[\frac{\hbar\Delta_c}{2}\pm 2\sqrt{3}\Omega^\prime\hbar\alpha]\sigma_z,
\end{aligned}
\label{hzpm}
\end{equation}
where $\alpha=-\sum_{l=1}^3\sin(\phi_{l+1}-\phi_l)$.

\begin{figure} 
\epsfig{figure=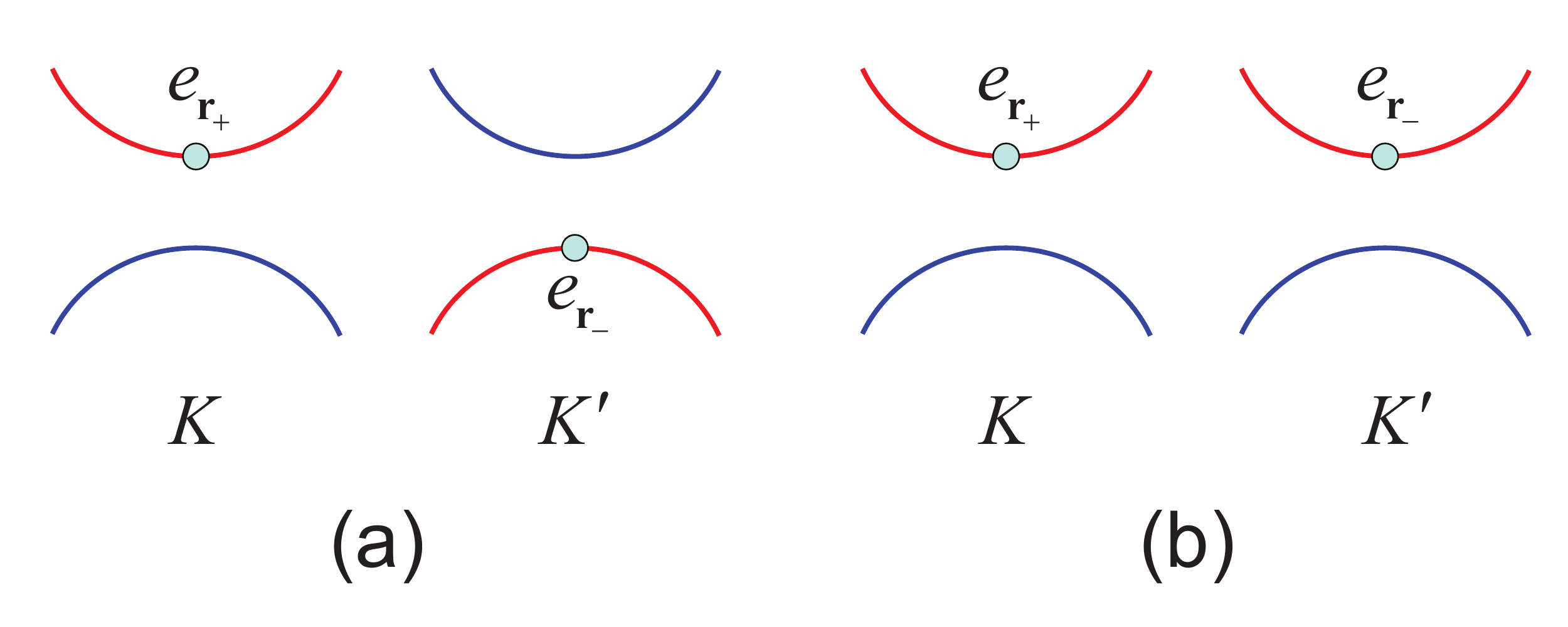, angle=0, width=0.45\textwidth}   
  \caption{{ The eigenstates at $K$ and $K^\prime$ points for topological and trivial SL's.} (a) For a topological SL, the eigenstate $|e\ra$ at $\mathbf{r}_+$ (i.e., $K$ point) is in the upper band while it is in the lower band at $\mathbf{r}_-$ (i.e., $K^\prime$ point). (b) For a trivial SL, $|e\ra$ is the eigenstate in the upper band at both $\mathbf{r}_+$ and $\mathbf{r}_-$. Red (Blue) lines denote the bands with more components in $|e\ra$ ($|m\ra$).} 
  \label{kk} 
\end{figure}

Let us consider the specific case that $\Delta_c=0$, $\phi_l=(l-1)4\pi/3$, and thus $\alpha=3\sqrt{3}/2$. In this case $\mathcal{H}(\mathbf{r}_\pm)=\pm 9\hbar\Omega^\prime\sigma_z$ has opposite signs at $\mathbf{r}_\pm$, as shown in Fig.\ref{effb} (a). It is obvious that at $\mathbf{r}_+$, the eigenstate in the upperband with eigenenergy $9\hbar\Omega^\prime$ is $|e\ra$, while at $\mathbf{r}_-$, $|e\ra$ is the eigenstate in the lower band, as shown in Fig.\ref{kk} (a). 

Near $\mathbf{r}_+$, we can use equation ({\ref{psi1}}) to describe the wavefunction. However, equation ({\ref{psi1}}) cannot describe all the wavefunctions in the upper band. It has a singularity at $\mathbf{r}_-$ where the magnetic field ${\mathbf h}$ points to the south pole.
We can remove the singularity by tuning $\Delta_c$. When $\Delta_c>4\sqrt{3}\Omega^\prime\alpha$, the effective magnetic field $\mathbf{h}$ does not experience the whole Bloch sphere [see Fig.\ref{effb} (b)]. In this case, $h_z$ in equation (\ref{hzpm}) has the same sign for $\mathbf{r}_\pm$, $|e\ra$ is in the upper band at both $\mathbf{r}_\pm$, as shown in Fig.\ref{kk} (b). The wavefunction can be described by the same gauge. Thus the single-valued Berry connection $\mathbf{A}$ has no singularities on a closed surface and the Chern number is zero. Generally, the Chern number of the upper band can be written as 
\begin{equation}
C=\frac{1}{2}\left[\text{sign}\left(\Delta_c+4\sqrt{3}\Omega^\prime\alpha\right)-\text{sign}\left(\Delta_c-4\sqrt{3}\Omega^\prime\alpha\right)\right].
\end{equation}
Specifically, when $\Delta_c=0$, $C=\text{sign}(\alpha)$.

\section{Dynamic evolution}
\label{sc5}

We can solve the dynamics of the atomic ensemble in real space with the total Hamiltonian including the probe field,
\begin{equation}
\begin{aligned}
H_t=&\sum\limits_{j=1}^{N}(h_z-\hbar\tilde{\nu}_p)|e_{\mathbf{r}_j}\ra \la e_{\mathbf{r}_j}|-(h_z+\hbar\tilde{\nu}_p)|m_{\mathbf{r}_j}\ra | m_{\mathbf{r}_j}|\\
&+[(h_x-ih_y)|e_{\mathbf{r}_j}\ra \la m_{\mathbf{r}_j}|+h.c.]\\
&-[\hbar\Omega_p e^{i\mathbf{k}_p\cdot\mathbf{r}_j}|e_{\mathbf{r}_j}\ra \la G|+h.c.],
\end{aligned}
\end{equation}
with $\tilde{\nu}_p=\Delta_c/2-\Delta_p$ being the probe detuning with respect to the middle of the band gap and $\Omega_p$ the probe field Rabi frequency.
The wavefunction in real space
\begin{equation}
|\Psi\ra=c_G|G\ra+\sum\limits_{j=1}^{N}c_e({\mathbf{r}_j})|e_{\mathbf{r}_j}\ra+c_m({\mathbf{r}_j})|m_{\mathbf{r}_j}\ra.
\end{equation}
The dynamic equations of the probability amplitudes are
\begin{equation}
\begin{aligned}
\dot{c}_e(\mathbf{r}_j)=&\left[-\frac{i}{\hbar}(h_z-\hbar\tilde{\nu}_p)-\gamma_e\right]c_e(\mathbf{r}_j)\\
&-\frac{i}{\hbar}(h_x-ih_y)c_m(\mathbf{r}_j)+i\Omega_p e^{i\mathbf{k}_p\cdot\mathbf{r}_j}c_G,
\end{aligned}
\label{ce}
\end{equation}
\begin{equation}
\begin{aligned}
\dot{c}_m(\mathbf{r}_j)=&\left[\frac{i}{\hbar}(h_z+\hbar\tilde{\nu}_p)-\gamma_m\right]c_m(\mathbf{r}_j)\\
&-\frac{i}{\hbar}(h_x+ih_y)c_e(\mathbf{r}_j).
\end{aligned}
\label{cm}
\end{equation}
where $\gamma_{e/m}$ is the decoherence rate of $|e/m\ra$ states. 

In the limit of weak probe field $\Omega_p\ll\gamma_e$, we have $c_G\approx 1$. In the steady state, $\dot{c}_e(\mathbf{r}_j)=\dot{c}_m(\mathbf{r}_j)=0$. From Eq.(\ref{cm}), we obtain
\begin{equation}
c_m(\mathbf{r}_j)=\frac{h_x+ih_y}{h_z+\hbar\tilde{\nu}_p+i\hbar\gamma_m}c_e(\mathbf{r}_j).
\label{cem}
\end{equation}
Substituting Eq.(\ref{cem}) in Eq.(\ref{ce}), we obtain
\begin{equation}
c_e(\mathbf{r}_j)=\frac{\hbar\Omega_p e^{i\mathbf{k}_p\cdot\mathbf{r}_j}}{h_z-\hbar\tilde{\nu}_p-i\hbar\gamma_e+\frac{h_x^2+h_y^2}{h_z+\hbar\tilde{\nu}_p+i\hbar\gamma_m}}.
\label{cr}
\end{equation}

In the SL coordinates, the wavefunction can be written as
\begin{equation}
|\Psi\ra=c_G|G\ra+\sum\limits_{\mathbf{k}}c_\mathbf{k}|e_\mathbf{k}\ra+c_{\mathbf{k}-\mathbf{k}_1}|m_{\mathbf{k}-\mathbf{k}_1}\ra.
\end{equation}
where the probability amplitude
\begin{equation}
c_\mathbf{k}=\la e_\mathbf{k}|\Psi\ra=\frac{1}{\sqrt{N}}\sum\limits_{j=1}^{N}e^{-i\mathbf{k}\cdot\mathbf{r}_j}c_e(\mathbf{r}_j).
\end{equation}
For uniformly distributed atoms, in the limit $N\gg\Omega_s/\gamma_e$, we can assume all points in the Brillouin zone are occupied by atoms, so the summation can be written as integration in the first Brillouin zone,
\begin{equation}
c_\mathbf{k}=\frac{\sqrt{N}}{S}\oiint_{\text{BZ}}c_e(\mathbf{r})e^{-i\mathbf{k}\cdot\mathbf{r}}\,dx\,dy,
\label{ck}
\end{equation}
where $S$ is the area of the first Brillouin zone.

\section{Diffraction Contrast and topology}

In this section, we quantitatively compare the contrast between $|c_{\mathbf{k}_+}|^2$ and $|c_{\mathbf{k}_-}|^2$ in topological and trivial SL's. We assume $\tilde{\nu}_p=0$ and $\gamma_m=0$. Then Eq.(\ref{cr}) becomes
\begin{equation}
c_e(\mathbf{r})=\frac{ h_z}{h^2-i\hbar h_z\gamma_e}\hbar\Omega_p e^{i\mathbf{k}_p\cdot\mathbf{r}}.
\end{equation}
We assume $\Omega^\prime,\Delta_c\ll\Omega_s$ and thus $c_e(\mathbf{r})$ is highly centred at $K$ and $K^\prime$ points, where we have $h_x=h_y=0$ and
\begin{equation}
c_e(\mathbf{r})=\frac{\hbar\Omega_p}{h_z-i\hbar\gamma_e} e^{i\mathbf{k}_p\cdot\mathbf{r}}.
\end{equation}
According to Eq.(\ref{ck}), the probability amplitude $c_\mathbf{k}$ can be approximately calculated by the integration of $c_e(\mathbf{r})$ in small areas near $\mathbf{r}_\pm$,
\begin{equation}
\begin{aligned}
c_\mathbf{k}\approx&\frac{\sqrt{N}\hbar\Omega_p}{S}[e^{i(\mathbf{k}_p-\mathbf{k})\cdot\mathbf{r}_+}\iint\limits_{\mathbf{r}_+}(h_z-i\hbar\gamma_e)^{-1}\,dx\,dy\\
&+e^{i(\mathbf{k}_p-\mathbf{k})\cdot\mathbf{r}_-}\iint\limits_{\mathbf{r}_-}(h_z-i\hbar\gamma_e)^{-1}\,dx\,dy],
\end{aligned}
\end{equation}
where the phase factors have been taken out of the integration since they do not change appreciably in those small areas.

For a topological SL where $\Omega^\prime\ne 0$ and $\Delta_c=0$, $h_z$ has opposite signs at $\mathbf{r}_\pm$. We denote $\iint_{\mathbf{r}_+}(h_z-i\hbar\gamma_e)^{-1}\,dx\,dy=p+iq$, then $\iint_{\mathbf{r}_-}(h_z-i\hbar\gamma_e)^{-1}\,dx\,dy=-p+iq$. Thus
\begin{equation}
\begin{aligned}
c_\mathbf{k}\approx 2i\frac{\sqrt{N}\hbar\Omega_p}{S}\{p\sin[{(\mathbf{k}_p-\mathbf{k})\cdot\mathbf{r}_+}]+q\cos[{(\mathbf{k}_p-\mathbf{k})\cdot\mathbf{r}_+}]\}.
\end{aligned}
\end{equation}
For $\mathbf{k}_p=-\mathbf{k}_1$, $\mathbf{k}_+=-\mathbf{k}_3$ and $\mathbf{k}_-=-\mathbf{k}_2$, we have
\begin{equation}
\begin{aligned}
\eta\equiv\frac{|c_{\mathbf{k}_+}|^2-|c_{\mathbf{k}_-}|^2}{|c_{\mathbf{k}_+}|^2+|c_{\mathbf{k}_-}|^2}\approx\frac{2\sqrt{3}pq}{3p^2+q^2}.
\end{aligned}
\end{equation}

For a trivial SL where $\Omega^\prime= 0$ and $\Delta_c\ne 0$, $h_z$ has the same sign at $\mathbf{r}_\pm$. We have $\iint_{\mathbf{r}_+}(h_z-i\hbar\gamma_e)^{-1}\,dx\,dy=\iint_{\mathbf{r}_-}(h_z-i\hbar\gamma_e)^{-1}\,dx\,dy=p+iq$ and
\begin{equation}
\begin{aligned}
c_\mathbf{k}\approx 2\frac{\sqrt{N}\hbar\Omega_p}{S}(p+iq)\cos[{(\mathbf{k}_p-\mathbf{k})\cdot\mathbf{r}_+}].
\end{aligned}
\end{equation}
Therefore, $c_{\mathbf{k}_+}=c_{\mathbf{k}_-}$ and $\eta=0$.

For a trivial SL with $\Omega^\prime\ne 0$ and $\Delta_c\ne 0$, it is difficult to analytically calculate the results. Nevertheless, as shown in Fig.3 (b) in the main text, $\eta$ reduces to a small value in the $C=0$ phases especially for large $\Omega^\prime$.

\section{Coupled- wave equations}
\label{sc9}

In this section, we calculate the propagation of a weak probe pulse. We set $\mathbf{k}_p=-\mathbf{k}_1$, $\mathbf{k}_+=-\mathbf{k}_3$ and $\mathbf{k}_-=-\mathbf{k}_2$. There will be directional emission in $\mathbf{k}_\pm$ by TDS $|e_{\mathbf{k}_\pm}\rangle$. The emitted photons in $\mathbf{k}_\pm$ interact with the atomic ensemble the same way as the probe field $\mathbf{k}_p$, creating TDS and resulting in directional emission in the other two modes. For example, emission at ${\mathbf k}_+$ excites the atom ensemble and leads to emission at ${\mathbf k}_-$ and ${\mathbf k}_{p}$. Therefore, the three optical fields in $\mathbf{k}_{p,\pm}$ are coupled via the atoms. In the following, we derive the coupling equations.

The expectation value of the dipole moment of the $j$th atom at position $\mathbf{r}_j$ is
\begin{equation}
\begin{aligned}
\la\Psi|\hat{\mu}_j|\Psi\ra=&\sum\limits_{\mathbf{k}}c_\mathbf{k}\la G|\hat{\mu}_j|e_\mathbf{k}\ra+c.c.\\
=&\frac{\mu}{\sqrt{N}}\sum\limits_{\mathbf{k}}{c}_\mathbf{k}e^{i\mathbf{k}\cdot\mathbf{r}_j}+c.c.,
\label{plr}
\end{aligned}
\end{equation}
where $\hat{\mu}_j$ is the dipole operator of the $j$th atom and $\mu=\la g_j|\hat{\mu}_j|e_j\ra$. Since the atoms are homogeneously distributed, the polarization density as a function of positions is,
\begin{equation}
\begin{aligned}
P(\mathbf{r})=&\frac{N}{V}\frac{\mu}{\sqrt{N}}\sum\limits_{\mathbf{k}}{c}_\mathbf{k}e^{i\mathbf{k}\cdot\mathbf{r}}+c.c.,
\end{aligned}
\end{equation}
where $V$ is the volume of the atomic ensemble. The probe field Rabi frequency is $\Omega_p=\mu E_p/\hbar$ with $E_p$ being the probe field strength.
The polarization density contains the Fourier components in $\mathbf{k}_\pm$, $P_{\pm, p}=\epsilon_0\chi_\pm E_p$ with $\chi_\pm=\sqrt{N} \mu{c}_{\mathbf{k}_\pm}/V\epsilon_0E_p$ and $\epsilon_0$ is the permittivity in vacuum. The notation $P_{\pm, p}$ means the $\mathbf{k}_\pm$ Fourier component of polarization density generated by optical field in $\mathbf{k}_p$. Once modes $\mathbf{k}_\pm$ are excited and the corresponding fields $E_\pm$ are generated, they also polarize the atoms. For $\phi_1=0$, $\phi_2=4\pi/3$ and $\phi_3=2\pi/3$, the three fields have a cyclic relation between each other. We have $P_{p,+}=\epsilon_0\chi_- E_+$, $P_{-,+}=\epsilon_0\chi_+ E_+$, $P_{p,-}=\epsilon_0\chi_+ E_-$ and $P_{+,-}=\epsilon_0\chi_- E_-$.

We assume the quasi-static approximation in which the atoms are assumed in steady sates, i.e., $c_{\mathbf k}$ is given by equation (\ref{ck}). This approximation is justified when the pulse duration is much longer than the decoherence time $1/\gamma_e$. We also use the slowly-varying-envelop approximation. The fields in the three relevant modes $\mathbf{k}_{j}$ with $j=p,+$ and $-$ are denoted as $E_{j}(\mathbf{r})e^{-i\nu_pt+i\mathbf{k}_{j}\cdot\mathbf{r}}$ where $|\mathbf{k}_{j}|=\nu_p/c\equiv k_p$. The coupled-wave Maxwell equations for the three modes are
\begin{equation}
\left(\hat{\mathbf{k}}_p\cdot\nabla+\frac{1}{v_g}\frac{\partial}{\partial t}\right)E_p=\frac{ik_p}{2}\left(\chi_0 E_p+\chi_-E_++\chi_+E_-\right),
\end{equation}
\begin{equation}
\left(\hat{\mathbf{k}}_+\cdot\nabla+\frac{1}{v_g}\frac{\partial}{\partial t}\right)E_+=\frac{ik_p}{2}\left(\chi_0 E_+ +\chi_-E_-+\chi_+E_p\right),
\end{equation}
\begin{equation}
\left(\hat{\mathbf{k}}_-\cdot\nabla+\frac{1}{v_g}\frac{\partial}{\partial t}\right)E_-=\frac{ik_p}{2}\left(\chi_0 E_-+\chi_-E_p+\chi_+E_+\right),
\end{equation}
where $\hat{\mathbf{k}}_{p/+/-}$ are the unit vectors in the directions of $\mathbf{k}_{p/+/-}$ and $v_g$ is the group velocity of the pulses.

\begin{figure} 
\epsfig{figure=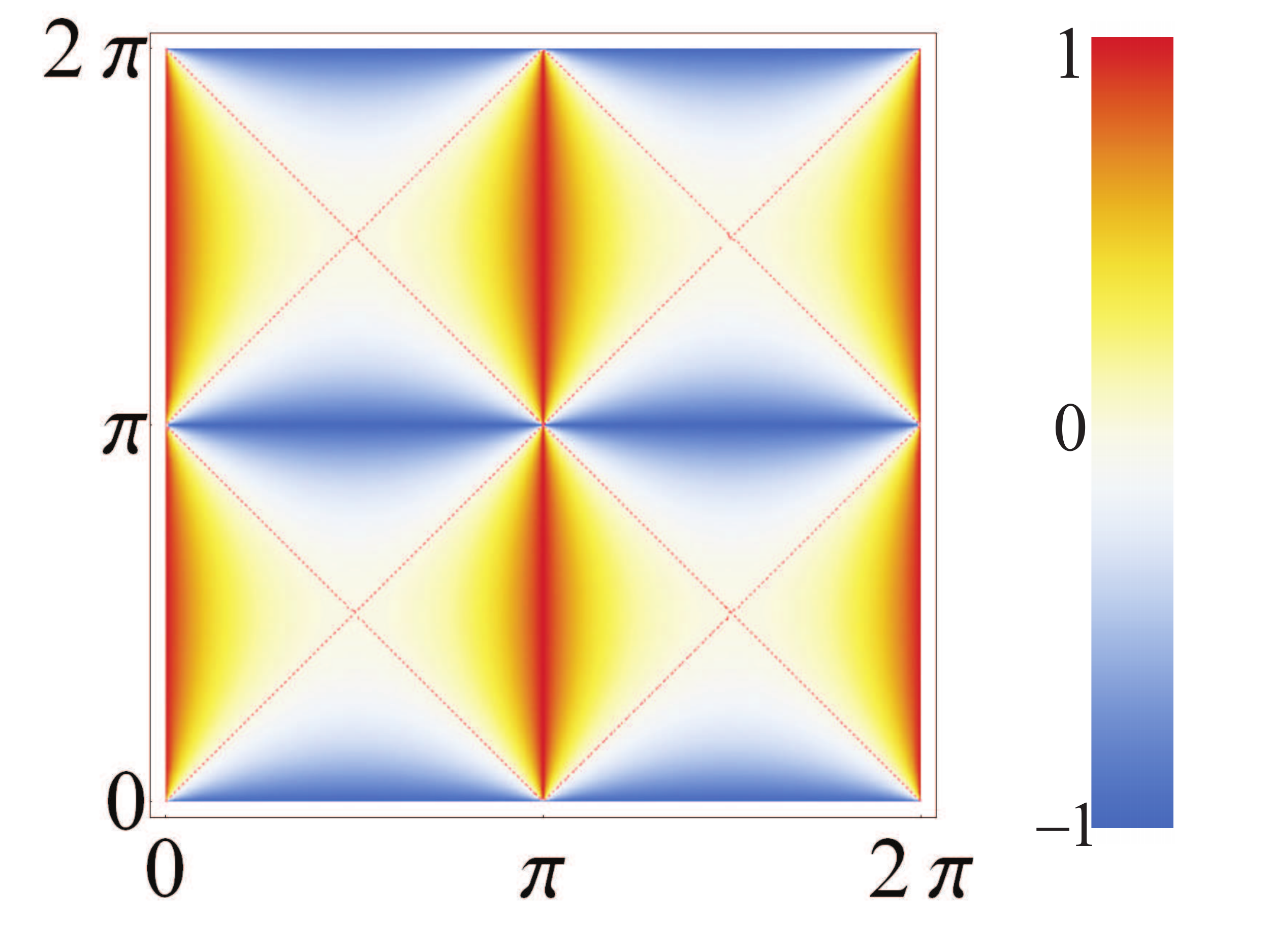, angle=0, width=0.4\textwidth}   
  \caption{{ The contrast of the NNN coupling strengths.} The contrast $(|\Omega_{13}|-|\Omega_{12}|)/(|\Omega_{13}|+|\Omega_{12}|)$ where $\Omega_{13}$ ($\Omega_{12}$) is the NNN hopping coefficient from $|e_{\mathbf{k}_p}\ra$ to $|e_{\mathbf{k}_+}\ra$ ($|e_{\mathbf{k}_-}\ra$). The dot lines are zero points.} 
  \label{cctrst} 
\end{figure}

\begin{figure} 
\epsfig{figure=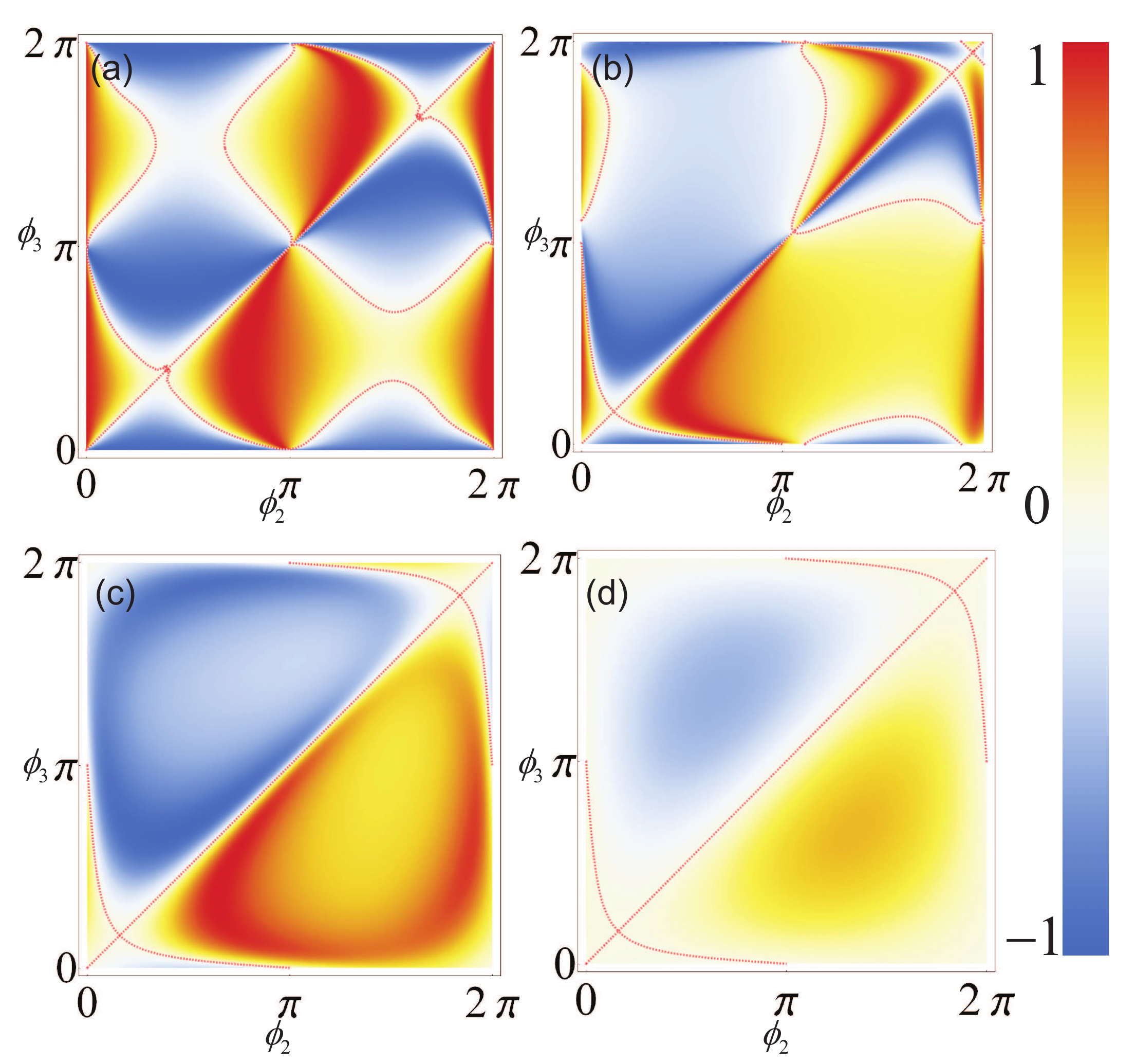, angle=0, width=0.5\textwidth}   
  \caption{{ The global and local effects.} Diffraction contrast $\eta$ for an SL with (a) $\Omega^\prime=1$, (b) $\Omega^\prime=0.01$, (c) $\Omega^\prime=0.001$ and (d) $\Omega^\prime=0.0001$. $\phi_1=0$. The dot lines are zero points. $\Omega_s=3$. $\Delta_p=\Delta_c=0$. $\gamma_e=1$ and $\gamma_m=0.001$.} 
  \label{s4} 
\end{figure}

\section{Local and global effects}
\label{sc11}

In the main text, we use the chiral topological current to explain the consistence between $\eta$ and the topological order. The topological current is a global effect for which we must treat the lattice as a whole. Since the three superradiant TDS $|e_{\mathbf{k}_{p,\pm}}\ra$ are next nearest neighbours in the lattice, the local effect induced by the transitions inside a unit cell is also important. The direct NNN transition strength from $|e_{\mathbf{k}_p}\ra$ to $|e_{\mathbf{k}_+}\ra$ or $|e_{\mathbf{k}_-}\ra$ is $\Omega_{13}$ or $\Omega_{12}$, respectively. If the dynamics is dominated by these direct transitions, we expect the contrast of the probability $\eta$ should be similar to $(|\Omega_{13}|-|\Omega_{12}|)/(|\Omega_{13}|+|\Omega_{12}|)$, which is plotted in Fig.\ref{cctrst}.

We plot the contrast $\eta$ in Fig.\ref{s4} for different $\Omega^\prime/\gamma_m$. The larger $\Omega^\prime$ is, the closer $\eta$ is to Fig.\ref{cctrst} and the more important is the local effect, as shown in Fig.\ref{s4} (a) and (b). For $\Omega^\prime\le\gamma_m$, the effect of the direct NNN transitions is suppressed by the decoherence rate $\gamma_m$. The global topological effect dominates for most areas and $\eta$ is closer to Fig.2 (a) in the main text, as shown Fig.\ref{s4} (c) and (d).

\end{document}